\documentclass[aps,twocolumn,showpacs,superscriptaddress,amsmath,amssymb,amsfonts,floatfix,longbibliography]{revtex4-1}

\usepackage{graphicx}  % Include figure files
\usepackage{multirow}
\usepackage{tabularx}
\usepackage{fancyhdr}
\usepackage{natbib,hyperref}
\usepackage{longtable}
\usepackage[T1]{fontenc}
\usepackage{dcolumn}   % Align table columns on decimal point
\usepackage{gensymb}
\usepackage{bm}        % bold math
\usepackage{amsfonts}  % Common math fonts
\usepackage{amsmath}   % Common math functions
\usepackage{amssymb}   % Common math symbols
\usepackage{hhline}
\usepackage{nicefrac}
\usepackage{ulem}
\newcommand{\mSR}{$\mu\rm{SR}$}

\hypersetup{
	colorlinks=true, %set true if you want colored links
	citecolor=blue,  %choose some color if you want links to stand out
}

\begin{document}

\title{Physical properties and electronic structure of the two-gap superconductor V$_{2}$Ga$_{5}$}

\author{P.-Y. Cheng}
\affiliation {Department of Physics, National Cheng Kung University, Tainan 701, Taiwan}
\affiliation {Center for Quantum Frontiers of Research \& Technology (QFort), National Cheng Kung University, Tainan 701, Taiwan}
\affiliation{Taiwan Consortium of Emergent Crystalline Materials, National Science and Technology Council, Taipei 10622, Taiwan}
\author{Mohamed Oudah}
\affiliation{Stewart Blusson Quantum Matter Institute, University of British Columbia, Vancouver, BC, V6T 1Z4, Canada}
\author{T.-L. Hung}
\affiliation {Institute of Physics, Academia Sinica, 128 Sec. 2, Academia Rd., Nankang, Taipei 11529, Taiwan}
\author{C.-E. Hsu}
\affiliation{\it Department of Physics, Tamkang University, New Taipei 251301, Taiwan}
\author{C.-C. Chang}
\affiliation {Department of Physics, National Cheng Kung University, Tainan 701, Taiwan}
\affiliation {Center for Quantum Frontiers of Research \& Technology (QFort), National Cheng Kung University, Tainan 701, Taiwan}
\affiliation{Taiwan Consortium of Emergent Crystalline Materials, National Science and Technology Council, Taipei 10622, Taiwan}
\affiliation {\it National Synchrotron Radiation Research Center (NSRRC), Hsinchu 30076, Taiwan}
\author{J.-Y. Haung}
\affiliation {\it Department of Physics, National Tsing Hua University, Hsinchu 30013, Taiwan}
\author{T.-C. Liu}
\affiliation {\it Department of Physics, National Tsing Hua University, Hsinchu 30013, Taiwan}
\author{C.-M. Cheng}
\affiliation {\it National Synchrotron Radiation Research Center (NSRRC), Hsinchu 30076, Taiwan}
\affiliation {\it Department of Physics, National Sun Yat-sen University, Kaohsiung 80424, Taiwan}
\affiliation {\it Department of Electrophysics, National Yang Ming Chiao Tung University, Hsinchu 300 Taiwan}
\affiliation {\it Taiwan Consortium of Emergent Crystalline Materials, National Science and Technology Council, Taipei 10622, Taiwan}
\author{M.-N. Ou}
\affiliation {Institute of Physics, Academia Sinica, 128 Sec. 2, Academia Rd., Nankang, Taipei 11529, Taiwan}
\author{W.-T. Chen}
\affiliation {Center for Condensed Matter Sciences, National Taiwan University, Taipei 10617, Taiwan}
\affiliation {Center of Atomic Initiative for New Materials, National Taiwan University, Taipei 10617, Taiwan}
\affiliation{Taiwan Consortium of Emergent Crystalline Materials, National Science and Technology Council, Taipei 10622, Taiwan}
\author{L. Z. Deng}
\affiliation {Texas Center for Superconductivity and Department of Physics, University of Houston, Houston, Texas 77204, United States}
\author{C.-C. Lee}
\affiliation{\it Department of Physics, Tamkang University, New Taipei 251301, Taiwan}
\author{Y.-Y. Chen}
\affiliation {Institute of Physics, Academia Sinica, 128 Sec. 2, Academia Rd., Nankang, Taipei 11529, Taiwan}
\author{C.-N. Kuo}
\affiliation {Department of Physics, National Cheng Kung University, Tainan 701, Taiwan}
\affiliation{Taiwan Consortium of Emergent Crystalline Materials, National Science and Technology Council, Taipei 10622, Taiwan}
\author{C.-S. Lue}
\affiliation {Department of Physics, National Cheng Kung University, Tainan 701, Taiwan}
\affiliation{Taiwan Consortium of Emergent Crystalline Materials, National Science and Technology Council, Taipei 10622, Taiwan}
\author{Janna Machts}
\affiliation{TRIUMF Centre for Molecular and Material Science (TRIUMF-CMMS), University of British Columbia, Vancouver, BC, V6T 2A3, Canada}
\affiliation{School of Physics and Astronomy, University of Edinburgh, Edinburgh EH9 3FD, United Kingdom}
\author{Kenji M. Kojima}
\affiliation{Stewart Blusson Quantum Matter Institute, University of British Columbia, Vancouver, BC, V6T 1Z4, Canada}
\affiliation{TRIUMF Centre for Molecular and Material Science (TRIUMF-CMMS), University of British Columbia, Vancouver, BC, V6T 2A3, Canada}
\author{Alannah M. Hallas}
\affiliation{Stewart Blusson Quantum Matter Institute, University of British Columbia, Vancouver, BC, V6T 1Z4, Canada}
\affiliation{Department of Physics \& Astronomy, University of British Columbia, Vancouver, BC, V6T 1Z1, Canada}
\affiliation{Canadian Institute for Advanced Research, Toronto, ON, M5G1M1, Canada}
\author{C.-L. Huang}
\email{clh@phys.ncku.edu.tw}
\affiliation {Department of Physics, National Cheng Kung University, Tainan 701, Taiwan}
\affiliation {Center for Quantum Frontiers of Research \& Technology (QFort), National Cheng Kung University, Tainan 701, Taiwan}
\affiliation{Taiwan Consortium of Emergent Crystalline Materials, National Science and Technology Council, Taipei 10622, Taiwan}

\date{\today}

\begin{abstract}  
We present a thorough investigation of the physical properties and superconductivity of the binary intermetallic V$_{2}$Ga$_{5}$. Electrical resistivity and specific heat measurements show that V$_{2}$Ga$_{5}$ enters its superconducting state below $T_{sc} =$ 3.5~K, with a critical field of $H_{\rm{c2},\perp c}(H_{\rm{c2},||c})$ = 6.5(4.1)~kOe. With $H\perp c$, the peak effect was observed in resistivity measurements, indicating the ultrahigh quality of the single crystal studied. The resistivity measurements under high pressure reveal that the $T_{sc}$ is suppressed linearly with pressure and reaches absolute zero around 20~GPa. Specific heat and muon spin relaxation measurements both indicate that the two-gap $s$-wave model best describes the superconductivity of V$_{2}$Ga$_{5}$. The spectra obtained from angle-resolved photoemission spectroscopy measurements suggest that two superconducting gaps open at the Fermi surface around the Z and $\Gamma$ points. These results are verified by first-principles band structure calculations. We therefore conclude that V$_{2}$Ga$_{5}$ is a phonon-mediated two-gap $s$-wave superconductor.
\end{abstract}

\maketitle

\section{Introduction}
Intermetallic compounds composed of metals of $d-$ and $p-$shell electrons have been one of the most studied series in condensed matter physics owing to their abundance and their chemical and physical properties. In particular, due to the low-melting points of metals with $p-$shell electrons, high-quality single crystals can often be grown by the common flux method and hence detailed property measurements can be conducted \cite{Canfield1992,Kanatzidis2005}.

\begin{figure*}
\includegraphics[width=.7\linewidth]{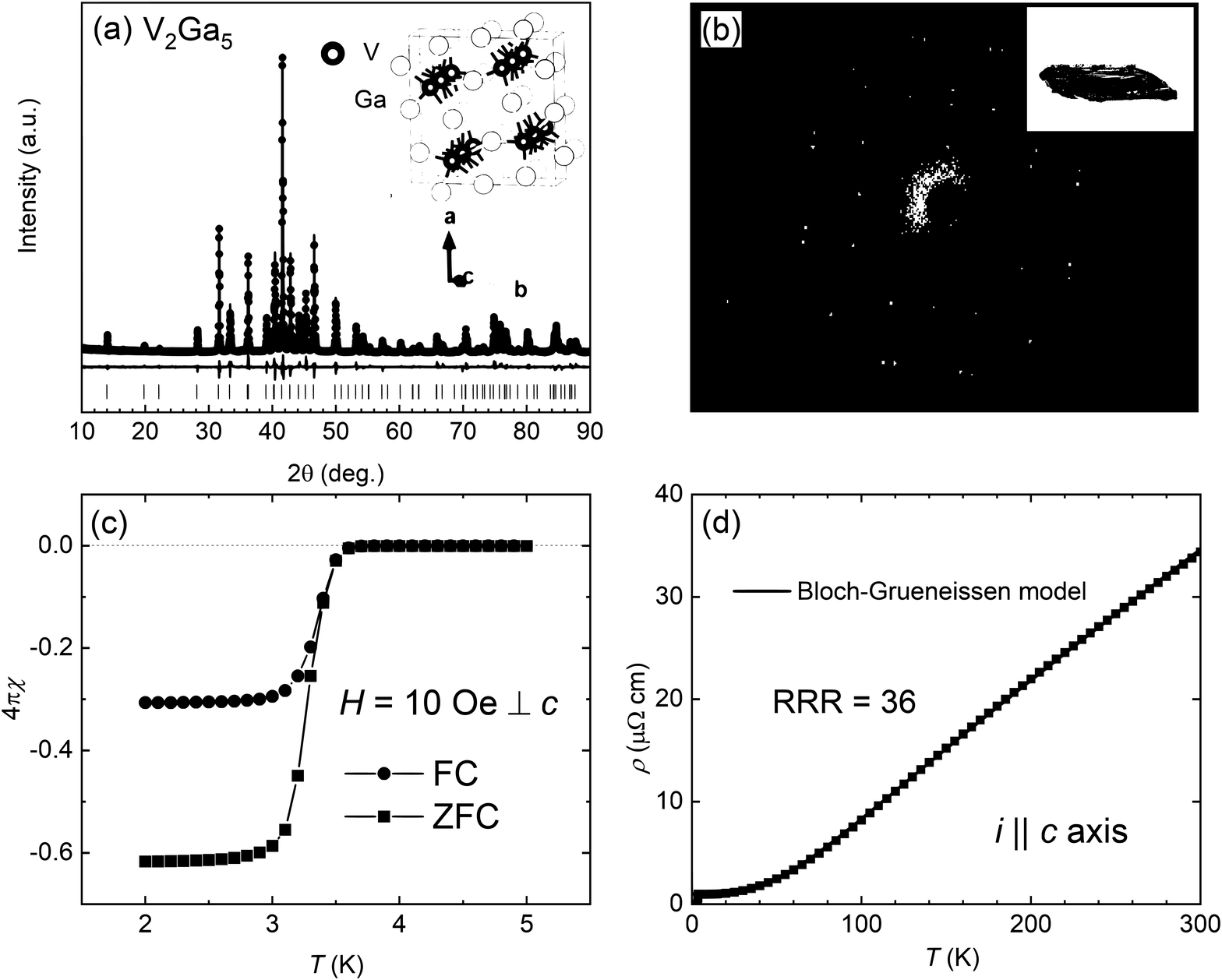}
\caption{(a) Observed (circle), calculated (red line), and differences (blue line) x-ray powder-diffraction patterns of V$_{2}$Ga$_{5}$ at room temperature. Bragg reflections for V$_{2}$Ga$_{5}$ of the space group $P$4/$mbm$     are labeled as vertical bars. The inset shows the crystal structure of one unit cell. Note that $a$ and $b$ are equivalent in tetragonal symmetry. (b) Laue x-ray pattern taken along the tetragonal [001] direction. The inset shows a photo of the crystal. (c) Field-cooled (FC) and zero-field-cooled (ZFC) magnetic susceptibility measurements at $H = 10$ Oe applied perpendicular to the $c$ axis. (d) Zero-field temperature dependence of electrical resistivity with a fit by the Bloch-Grueneissen model.}  
\label{Fig1} 
\end{figure*}

Among mentioned intermetallic compounds, the V-Ga binary system has gained extra interest because some compounds in this family are superconductors. The superconductor V$_{3}$Ga is best known for its relatively high transition temperature ($T_{sc}$ = 15~K), high critical currents, and fields \cite{Goodman1962,Junod1971}. In contrast, for another superconductor V$_{2}$Ga$_{5}$, relevant research reports are quite scarce. The electrical resistivity, magnetic susceptibility, and specific heat measurements conducted down to 1.8~K have been reported, which showed V$_{2}$Ga$_{5}$ enters the superconducting state below $T_{sc}$ = 3.5~K and its critical field is about 4-5 kOe \cite{Cruceanu1974,Lobring2002,Teruya2015}. Very recently, Xu et al. report multigap superconductivity in V$_{2}$Ga$_{5}$ via specific heat and thermal conductivity measurements \cite{Xu2024}. To verify this result and acquire a deeper understanding of the superconductivity and the physical properties of V$_{2}$Ga$_{5}$, it is essential to expand the specific heat measurements to lower temperatures and conduct a comprehensive study using various experimental techniques. We carried out electrical resistivity measurements with magnetic fields applied parallel and perpendicular to the crystallographic $c$ axis and constructed the critical field vs. $T_{sc}$ phase diagrams. Under high-pressure, $T_{sc}$ is suppressed linearly to absolute zero around $P = 20$~GPa. This result is in line with the phonon mediated superconductivity in V$_{2}$Ga$_{5}$. A consistent conclusion about the two-gap $s$-wave order parameter is drawn from two complimentary experiments: specific heat as a bulk probe and muon spin relaxation as a local probe.  This conclusion was further validated by angle-resolved photoemission spectroscopy experiments and first-principles calculations.

\section{Research methods}
V$_{2}$Ga$_{5}$ single crystals were prepared by a flux method using excess Ga \cite{Lobring2002,Teruya2015}. High-purity V powder and Ga pieces were weighed according to the composition V$_{6}$Ga$_{94}$ and placed in an alumina crucible, which was then sealed in an evacuated quartz tube. Higher vanadium concentrations were also tested but yielded needle-shape samples, as been reported in Ref.~\cite{Cruceanu1974,Lobring2002,Teruya2015}. The quartz tube was heated to 1000~\degree C at a rate of 100~\degree C/h, held at this temperature for 48~h and then cooled to 550~\degree C at a rate of 2.5~\degree C/h. Excess Ga flux was spun off using a centrifuge. The remaining Ga droplets on the surface were removed by etching in diluted HCl. As shown in the inset of Fig.~\ref{Fig1}(b), the obtained single crystals have rod-like shape along the c-axis with typical dimensions $\sim 2 \times 2 \times 8$ mm$^{3}$. Pulverized V$_{2}$Ga$_{5}$ was characterized by x-ray powder diffraction with Cu-K$\alpha$ radiation (Bruker D2 Phaser diffractometor) at room temperature. The crystal structure was analyzed using DIFFRAC.TOPAS program \cite{Coelho2018} with the Rietveld refinement method. Single crystal x-ray diffraction data were collected with a Rigaku Synergy-S diffractometer with Mo-K radiation at 300 and 100 K. Images of representative reciprocal planes are shown in Supplemental Material \cite{SM}. The orientation of crystal was confirmed by a Laue X-ray diffraction method. Reciprocal planes from single crystal x-ray diffraction and Laue diffractograms validate  the good quality of the single-crystalline samples and an example is shown in Fig.~\ref{Fig1}(b).

\begin{table}[]
\caption{Crystal data and structural refinement for V$_{2}$Ga$_{5}$. $a$ and $c$ are lattice parameters. $V$ is the unit-cell volume. $B_{iso}$ is the thermal parameter. $\chi^{2}$ represents the goodness of fit.  $R_{wp}$ and $R_{p}$ are weighted and unweighted profile $R$ factors, respectively, obtained from the refinement results. The gallium atom occupies Wyckoff positions Ga1 2$d$ (\nicefrac{1}{2}, 0, 0) and Ga2 8$i$ ($x$, $y$, 0), while vanadium atom occupies V 4$h$ ($x$, $x$+\nicefrac{1}{2}, \nicefrac{1}{2}) position.}
\begin{tabular}{ccccccc}
\hhline{=======}
\multicolumn{7}{c}{Tetragonal $P$4/$mbm$ (No. 127)}             \\
\multicolumn{7}{c}{$a$ = 8.9716(1) \r{A}, $c$ = 2.69272(4) \r{A}, $V$ = 216.737(7) \r{A}$^{3}$} \\ \\
Atom     & $x$             & $y$             & $z$          & Occupancy     & $B_{iso}$      & Site   \\ \hline
Ga1      & 0.2942(1)     & 0.5611(1)     & 0     & 1.000         & 0.02(4)   & 8$i$     \\
Ga2      & 0.5        & 0        & 0     & 1.000         & 0.02      & 2$d$     \\
V        & 0.3212(2)     & 0.8212(2)     & 0.5     & 1.000         & 0.09(6)   & 4$h$     \\ \\
\multicolumn{7}{c}{$\chi^{2}$ = 1.60, $R_{wp} = 11.36$ \% and $R_{p} = 8.73$ \%}   \\                           \hline
\hline
\end{tabular}
\label{Table 1}
\end{table}

Magnetic susceptibility was measured using a Quantum Design (QD) magnetic property measurement system. Ambient-pressure resistivity and specific heat were measured using a QD Dynacool physical properties measurement system (PPMS) equipped. For high-pressure resistivity measurements, we used a diamond anvil cell (DAC) placed in a $^{3}$He cryostat. For high-pressure Hall effect measurements, we used a DAC placed in QD Dynacool PPMS. 

Muon spin relaxation ($\mu$SR) measurements were performed on the M15 beam line at TRIUMF using the Pandora dilution refrigerator spectrometer. Several dozen needle-like crystals of V$_2$Ga$_5$ were co-aligned by eye on a silver sample plate using GE varnish. Measurements were performed over a temperature range of 0.025 K to 6 K in both zero field and in an applied transverse field of 300~Oe. To achieve an accurate zero field, reference measurements were performed on silicon, which through the formation of muonium atoms with a two orders of magnitude larger gyromagnetic ratio than the muon, allows the magnetic field to be zeroed to within 10$^{-2}$~Oe~\cite{morris2003method}. In the zero field measurements, the muons were implanted with their spins parallel to the $ab$-plane of  V$_2$Ga$_5$ while in the transverse field measurements, the muons are implanted with their spins along the $c$-axis and the transverse field is applied parallel to the $ab$-plane. The collected data were analyzed with \texttt{MUSRFIT}~\cite{suter2012musrfit}.

Angle-resolved Photoemission Spectroscopy (ARPES) experiment was conducted at beamline BL21B1 of the Taiwan Light Source, affiliated with the National Synchrotron Radiation Research Center. The whole experiment data were collected by Scienta R4000 analyzer within ultra-high vacuum (UHV) condition, utilizing an angular resolution calibrated to 0.5 degrees. Single crystal of V$_2$Ga$_5$ was cleaved in the preparation chamber maintained under UHV conditions to ensure the freshness of the crystal plane, with a pressure of approximately 8.9×$10^{-11}$ torr. The spectra were obtained at 89 K, using photon energies between 52 eV and 86 eV, with the measurements achieving an overall energy resolution exceeding 22 meV.

The first-principles calculations were performed using Quantum Espresso \cite{Giannozzi_2017} code with PBE exchange-correlation functional. The atomic positions were optimized at the experimental lattice constants. The plane wave cutoff was set to 100 Ry and the k points were sampled on a 4$\times$4$\times$12 grid. The electronic structure was calculated with the inclusion of spin-orbit coupling.
For the electron-phonon properties, the EPW \cite{Ponce2016} software was used to interpolate the electron-phonon coupling matrix elements from the coarse grid (4$\times$4$\times$12 k and 2$\times$2$\times$6 q grids) to the finer grid (20$\times$20$\times$60 k and 10$\times$10$\times$30 q grids). For the computational efficiency, the effect of spin-orbit coupling has been neglected when 
calculating the electron-phonon-related properties using the linear response framework.
Note that the band structure obtained here without adopting any Hubbard correction is similar to the GGA+U result with U=1 eV, where the band topology at the Fermi level is consistent with the result of U = 3.5~eV \cite{Xu2024}.

\section{Results and discussions}
\subsection{Crystal structure}
The refined structural parameters are summarized in Table~\ref{Table 1}. It was shown that the crystal is single phase and crystallized in tetragonal space group $P$4/$mbm$ (No. 127), which is in consistent with reported crystal structure \cite{Vucht1964}. The crystal structure can be viewed as connecting and piling of slightly distorted vanadium-centered VGa$_{10}$ pentagonal pillars, in which consisting of eight longer V-Ga1 bonds of 2.72~\r{A} and two shorter V-Ga2 bonds of 2.59~\r{A}.

\subsection{Electrical resistivity}
As V$_{2}$Ga$_{5}$ enters the superconducting state below $T \sim 3.5$ K, a diamagnetic signal is clearly seen in magnetization measurements, as shown in Fig.~\ref{Fig1}(c). The demagnetization effect has been taken into account \cite{Cullity2008}. Figure~\ref{Fig1}(d) shows the temperature dependence of electrical resistivity $\rho$ for current along the tetragonal $c$ axis under zero magnetic field. The residual resistivity $\rho_{0} \sim 1.0$~$\mu\Omega$~cm (the value at the temperature right above the superconducting transition) and the residual resistivity ratio RRR = $\rho_{300 \rm{K}}/\rho_{4 \rm{K}} = 36$ are both comparable with the previous result \cite{Teruya2015}. The normal state $\rho$ could be described by the Bloch-Grueneisen model,

\begin{equation}
    \rho =\rho_{0}+A(\frac{T}{\theta_{R}})^{n}\int_{0}^{\frac{\theta_{R}}{T}}\frac{x^{n}}{(e^{x}-1)(1-e^{-x})}dx,
\end{equation}
where $A$ is a constant related electron-phonon interactions, $\theta_{R}$ is the Debye temperature as obtained from resistivity measurements, and $n$ is an integer that depends upon the nature of interaction. The excellent fit results in $A = 88.0$~${\mu\Omega}\rm{cm}$ and $\theta_{R} = 369$~K. An exponent $n = 3$ can best describe the data, implying the resistance in V$_{2}$Ga$_{5}$ is mainly due to $s-d$ scattering, as is the case for transition metals \cite{Kittel}. 

\begin{figure*}
\includegraphics[width=.8\linewidth]{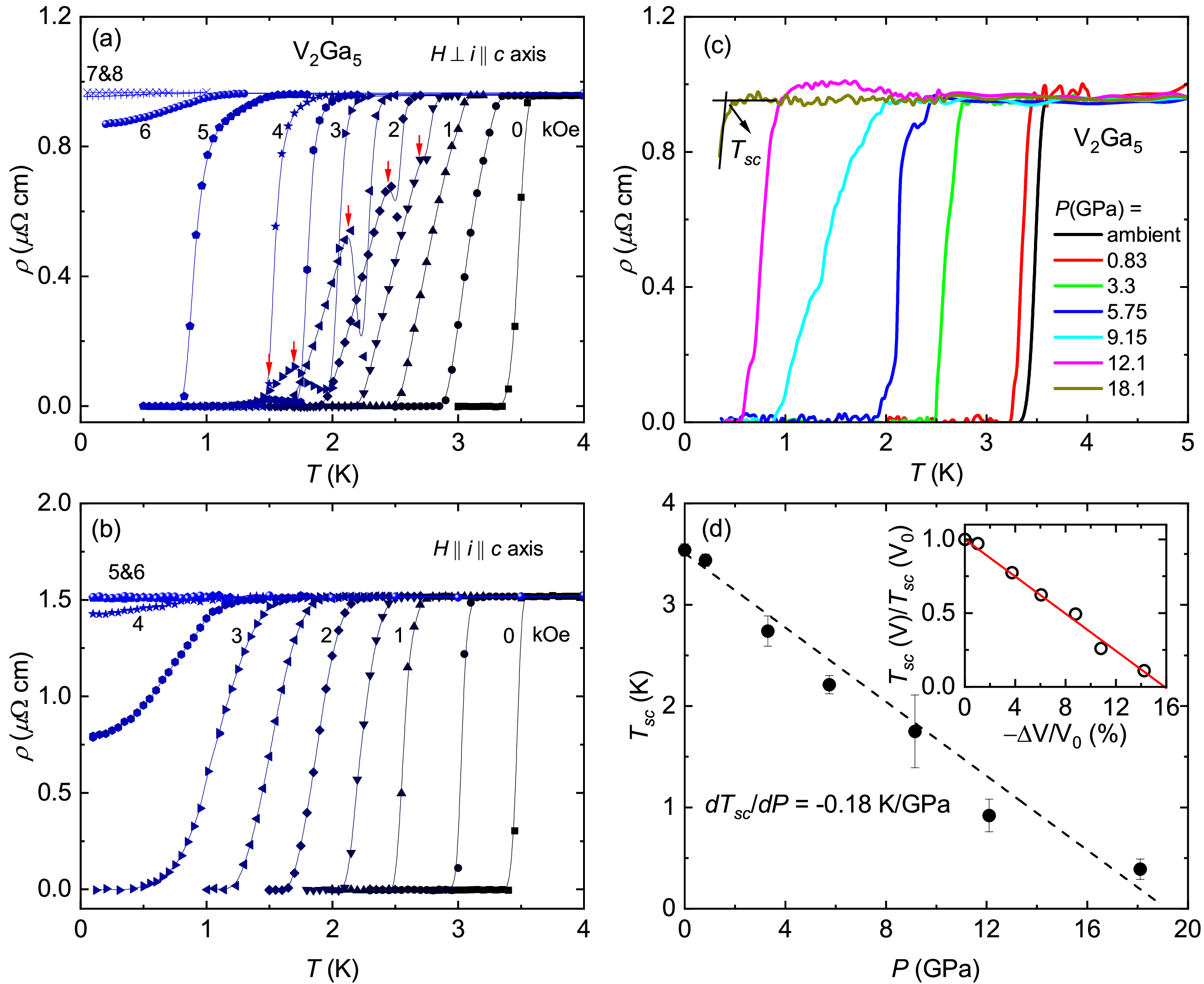}
\caption{Field and pressure dependence of the superconducting transition in resistivity of V$_{2}$Ga$_{5}$. (a) $\rho$ with $H$ applied perpendicular to the $c$ axis for V$_{2}$Ga$_{5}$. From right to left, $H = 0, 0.5, 1, 1.5, 2, 2.5, 3, 3.5, 4, 5, 6, 7,$ and 8~kOe. The downward red arrows point out the local resistivity maximum. (b) $\rho$ with $H$ applied parallel to the $c$ axis. From right to left, $H = 0, 0.5, 1, 1.5, 2, 2.5, 3, 3.5, 4, 5,$ and 6~kOe. Current $i =$ 0.1 mA is applied along the $c$ axis. (c) Temperature dependence of resistivity at different pressures $P$ for V$_{2}$Ga$_{5}$. The applied current is along the $c$ axis. An arrow demonstrates how $T_{SC}$ was determined. (d) $T_{sc}$ vs. $P$. The inset shows the relative pressure shift of $T_{sc}$ as function of the relative volume change. The dashed and solid lines are linear fits through data points.}  
\label{RT} 
\end{figure*}

Next we turn to magnetic field effect on low-$T$ resistivity. Figure~\ref{RT}(a) and (b) show the data measured with the magnetic field applied perpendicular to and parallel to the $c$ axis, respectively. Two different samples were used. The difference in normal state $\rho$ values between the two field directions is mostly attributed to errors in measuring the dimensions of the samples. The superconducting transition temperature $T_{sc}$ was determined as the temperature where two linear fits above and below the resistivity drop intersect, as shown by an arrow in Fig.~\ref{RT}(c). The good quality of our samples is confirmed by a sharp transition into the superconducting state at zero field. Upon application of a magnetic field, the transition broadens and moves to lower temperatures as expected. The critical field of $H_{c2,\perp c}(H_{c2,|| c})$ = 6.5 (4.1) kOe is much smaller than the Pauli limited value ($H_{c2} = 1.84 T_{sc} = 6.4$ T) and comparable to the value reported in the literature \cite{Xu2024}. Interestingly, $\rho$ shows the peak effect, which has been dubbed a rise in $\rho$ for a certain range of magnetic fields below the onset of the superconducting transition \cite{Tomy1997}, with $H\perp c$ axis, as shown in Fig.~\ref{RT}(a). This sudden increase in $\rho$ is observable for $H$ between 1.5 and 3.5~kOe at the measured current density of $\sim$1 A/cm$^{2}$, and is reversible for increasing and decreasing the temperatures. The peak effect has been observed in several single crystal superconductors with two- or three-dimensional crystal structures \cite{Autler1962,Higgins1996,Tomy1997,Hedo1998,Chaudhary2001,Lee2008,Kaluarachchi2016}. The peak effect may be attributed to the absence of pinning centers in the sample, indicating high sample quality. This allows the vortex to flow around the sample, thereby forming a route for the current to dissipate energy \cite{Willa2018}. We note that this is the first time the peak effect is observed in V$_{2}$Ga$_{5}$. It is absent in the data reported by Xu et al. where the sample has a higher residual resistivity \cite{Xu2024}.

\begin{table}[]
\setlength{\tabcolsep}{0.2\tabcolsep}
\caption{Superconducting parameters derived from current work and the literature. $\xi$ stands for the coherence length of superconducting Cooper pairs.}
\begin{tabularx}{\columnwidth}{ccccc}
\hhline{=====}
parameters & Ref.~\cite{Lobring2002} &  Ref.~\cite{Teruya2015} & Ref.~\cite{Xu2024}  & present work   \\ \hline
$T_{sc}$ & 3.59  & 3.3 & $3.6$  & 3.5   \\
$H_{c2,\perp c}(0)$(kOe) & 3-4 & $-$ & 5.7 & 6.5 \\
$H_{c2,||c}(0)$(kOe) & $-$ & $-$ & 4.9 & 4.1 \\
$H_{c1,\perp c}(0)$(kOe) & $-$ & $-$ & $-$ & 0.036 \\
$H_{c1,||c}(0)$(kOe) & $-$ & $-$ & $-$ & 0.072\\
$dH_{c2,\perp c}/dT|_{T=T_{sc}}$(kOe/K) & $-$ & $-$ & $-$ & $-1.85$\\
$dH_{c2,|| c}/dT|_{T=T_{sc}}$(kOe/K) & $-$ & $-$ & $-$ & $-1.01$\\
$\xi_{ab}/\xi_{c}$ & $-$ & $-$ & $-$ & $1.83$\\
$\gamma_{n}$ (mJ mol K$^{2}$) & $-$ & 16.7$\pm0.2$ & 17.92 & 17.8$\pm0.1$ \\    
$\beta$ (mJ mol K$^{4}$) & $-$ & 0.448 & 0.024 & 0.20$\pm0.01$ \\
$\theta_{\rm D} (K)$ & $-$ & 312 & $-$ & 408 \\
$C_{e}/\gamma_{n}T_{sc}$ & $-$ & 0.64 & 0.92 & 1.0 \\
2$\Delta_{1}/k_{\rm B}T_{sc}$ & $-$ & $-$ & $-$ & 3.6 \\
2$\Delta_{2}/k_{\rm B}T_{sc}$ & $-$ & $-$ & $-$ & 1.5 \\
\hline
\end{tabularx}
\label{Table 2}
\end{table}
\subsection{Pressure dependence of superconductivity}
The pressure effect is a useful tool for varying the primary parameters that determine superconductivity, such as the density of states at the Fermi energy, the phonon frequency, and the coupling constant of electrons and phonons, without introducing chemical disorder. \cite{Lorenz2005}.  We investigated the pressure effect on superconductivity of V$_{2}$Ga$_{5}$ as shown in Fig.~\ref{RT}(c-d). Upon applying pressure, $T_{sc}$ decreases linearly at a rate of $-0.18$ K/GPa up to $P = 18.1$~GPa, from which we extrapolate $T_{sc} \rightarrow 0$ around $P = 20$ GPa.

In order to convert the applied pressure to volume compression, the bulk modulus $B \equiv V \times (dP/dV)$ needs to be known. Since there are no available experimental data of $B_{\rm {V_{2}Ga_{5}}}$, we estimate it using a Voigt-Reuss approximation method: from the bulk moduli of pure constituents $B_{i}$ with relative concentration $f_{i}$ in the compound, we calculate $B = (B_{Voigt} + B_{Reuss})/2$ with $B_{Voigt} = \sum f_{i}B_{i}$ and 1/$B_{Reuss} = \sum f_{i} /B_{i}$ yielding $B_{\rm {V_{2}Ga_{5}}} \sim 78.4$ GPa \cite{Grimvall1999,Kittel}. For a range of intermetallic compounds, this method yields $B$ values in agreement of with experimental data \cite{Man2011}. We then used a Murnaghan equation of state:
\begin{equation}
B = B_{0}+p\frac{dB}{dP}|_{p=0}
\end{equation}
and for cell volume one has:
\begin{equation}
\frac{V}{V_{0}}=\left [ 1+\frac{\frac{dB}{dP}|_{p=0}}{B_{0}}p \right ]^{-1/\frac{dB}{dP}|_{p=0}},
\end{equation}
where $B_{0}$ and $V_{0}$ are bulk modulus and cell volume at ambient pressure, respectively. For $\frac{dB}{dP}|_{p=0}$, an empirical value of 5 was used, which was found for most compounds \cite{Grube2022}. The obtained $T_{sc}$ as a function of relative volume change is shown in the inset of Fig.~\ref{RT}(d) where a linear relationship is observed. This behavior has been shown in a plethora of low temperature superconductors, which indicates phonon hardening under pressure, and hence suggest a phonon-mediated superconductivity for V$_{2}$Ga$_{5}$ \cite{Lorenz2005}.

\subsection{Specific heat}
\begin{figure}
\includegraphics[width=\columnwidth]{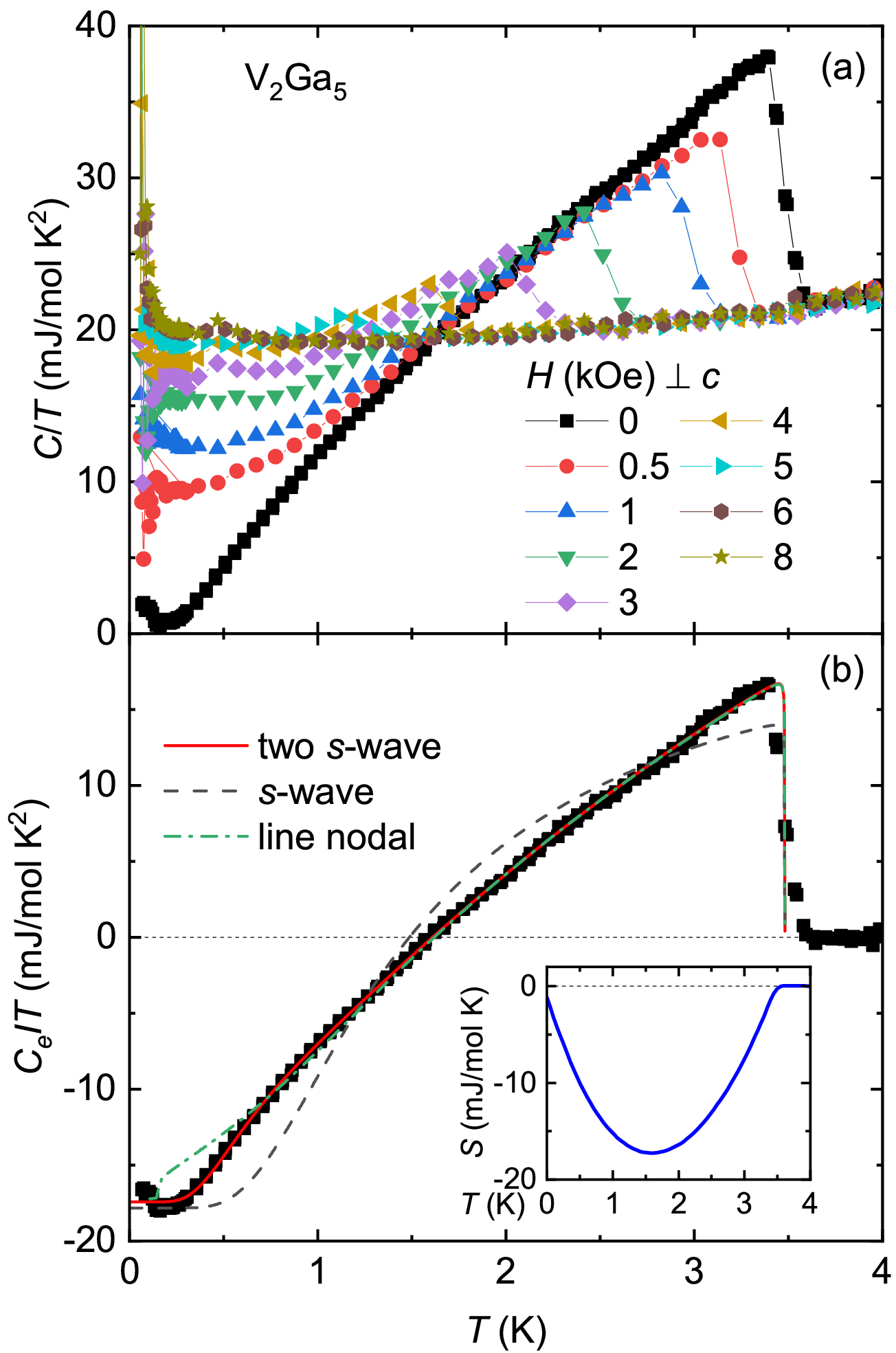}
\caption{(a) Temperature dependence of specific heat plotted as $C/T$ vs $T$ at various magnetic fields applied perpendicular to the $c$ axis for V$_{2}$Ga$_{5}$. (b) Temperature dependence of the superconducting-transition contribution to the specific heat $C_{e}/T$. The solid, dashed, and dash-dot lines are fits using a two-gap $s$-wave, a single-gap $s$-wave, and a line node $d$-wave models, respectively. Inset shows that the entropy ($S$), calculated from the data in the main panel, is conserved for the superconducting phase transition.}    
\label{CoverT} 
\end{figure}

Figure~\ref{CoverT}(a) shows the temperature dependence of the specific heat $C/T$ of V$_{2}$Ga$_{5}$ under different magnetic fields $H = 0$ to 8~kOe. In zero field, a jump in $C(T)$ representing the superconducting transition is observed at around 3.5~K, consistent with the finding in the previous specific-heat studies \cite{Teruya2015,Xu2024}. As the temperature decreases, $C/T$ approaches zero, indicating a complete superconducting volume of the sample. Below $T \sim 0.1$~K, an upturn of $C/T$ due to a nuclear Schottky contribution from the vanadium isotope is observed. This upturn occurs at higher temperatures with an increase of the magnetic field as expected \cite{Rai2018}. The superconducting transition can be traced up to $H = 5$~kOe and cannot be observed at $H \geq 6$~kOe. Thus, the normal-state specific heat $C_{n}(T)$ is extracted from the $H = 8$~kOe data between $T = 1$ to 4~K to avoid the nuclear contribution at lower temperatures. $C_{n}(T)$ can be described as $C_{n}(T) = \gamma_{n}T+\beta T^{3}$, where $\gamma_{n}T$ is the normal electronic contribution and $\beta T^{3}$ represents the phonon contribution. The fit (not shown) leads to $\gamma_{n} = 17.82\pm0.10$~mJ/mol~K$^{2}$ and $\beta = 0.200\pm0.012$~mJ/mol~K$^{4}$. From $\beta$, we derive the corresponding Debye temperature $\theta_{\rm D} = 408$~K, which is close to the value determined from our Bloch-Grueneissen fit to the resistivity, $\theta_{R} = 369$~K (Fig.~\ref{Fig1}(d)). The obtained $\gamma_{n}T$ value is consistent with that in Ref.~\cite{Xu2024}, but our $\beta$ value is almost one order of magnitude higher. This is most probably due to the fact that Xu et al. introduced a higher order term ($C_{n}(T) = \gamma_{n}T+\beta T^{3} + \delta T^{5}$) to describe the phonon contribution to the specific heat. Their fit yields an unphysically low $\theta_{\rm D}$, and we did not find a $T^{5}$ term necessary to describe the data.

The zero-field superconducting electronic contribution to the specific heat $C_{e}(T)$ is obtained with $C_{e} = C(H = 0)-C_{n}$, as shown in Fig.~\ref{CoverT}(b). The superconducting transition is a second-order phase transition which requires the entropy conservation. The entropy conservation is evidenced in the inset of Fig.~\ref{CoverT}(b). The dimensionless specific-heat jump $C_{e}/\gamma_{n}T_{sc} \sim 1.0$ at $T_{sc}$ is smaller than the weak coupling value 1.43 described in the BCS theory \cite{Bardeen1957}, and is comparable to the value reported in Ref.~\cite{Xu2024}. To examine the superconducting order parameter of V$_{2}$Ga$_{5}$, the data above the nuclear Schottky anomaly ($T > 0.2$~K) are fitted to a single-gap $s$-wave, a line-node with $d$-wave, and two-gap $s$-wave models with the following equation
\begin{equation}
\begin{split}
    C_{e}=2N(0)\beta k\frac{1}{4\pi}\int_{0}^{2\pi}d\phi\int_{0}^{\pi}d\theta \textup{sin}\theta\\
    \times \int_{-\hbar\omega_{D}}^{\hbar\omega_{D}}-\frac{\partial f}{\partial E}\left ( E^{2}+\frac{1}{2}\beta\frac{d\Delta^{2}}{d\beta} \right )d\varepsilon, 
\end{split}
\end{equation}
where $N(0)$ is the density of states at the Fermi surface, $\beta = 1/k_{\rm B}T$, $E = (\varepsilon^{2}+\Delta^{2})^{1/2}$, $f = (1+e^{\beta E})^{-1}$, $\Delta = \Delta_{0}$ the superconducting gap for an isotropic $s$-wave, $\Delta = \Delta_{0}$cos$2\phi$ for line nodes with $d$-wave symmetry, and $\Delta = \Delta_{1}\cdot x+\Delta_{2}\cdot(1-x)$ where $x$ is relative weighting for two-gap $s$-wave. From the fitting result, it is obvious the one-gap $s$-wave cannot describe the data. For the line-nodal $d-$wave model, the fit agrees well with the data between $T_{sc}$ and 0.5~K, but becomes unsatisfactory at lower temperatures. Finally, an excellent fit using a two-gap $s$-wave model with 2$\Delta_{1}/k_{\rm B}T_{sc} = 3.6$ and 2$\Delta_{2}/k_{\rm B}T_{sc} = 1.5$, with a relative weighting of 55\% and 45\%, respectively, across the entire fitting range is observed. The two superconducting gap values obtained from this fit are 0.543 meV and 0.226 meV. Therefore, we conclude V$_{2}$Ga$_{5}$ is a two-gap $s$-wave superconductor. This conclusion can only be drawn if specific heat measurements are conducted below 0.5 K. Above 0.5 K, both the two-gap $s$-wave and $d$-wave models can describe the data equally well \cite{Xu2024}.

\begin{figure}
\includegraphics[width=\columnwidth]{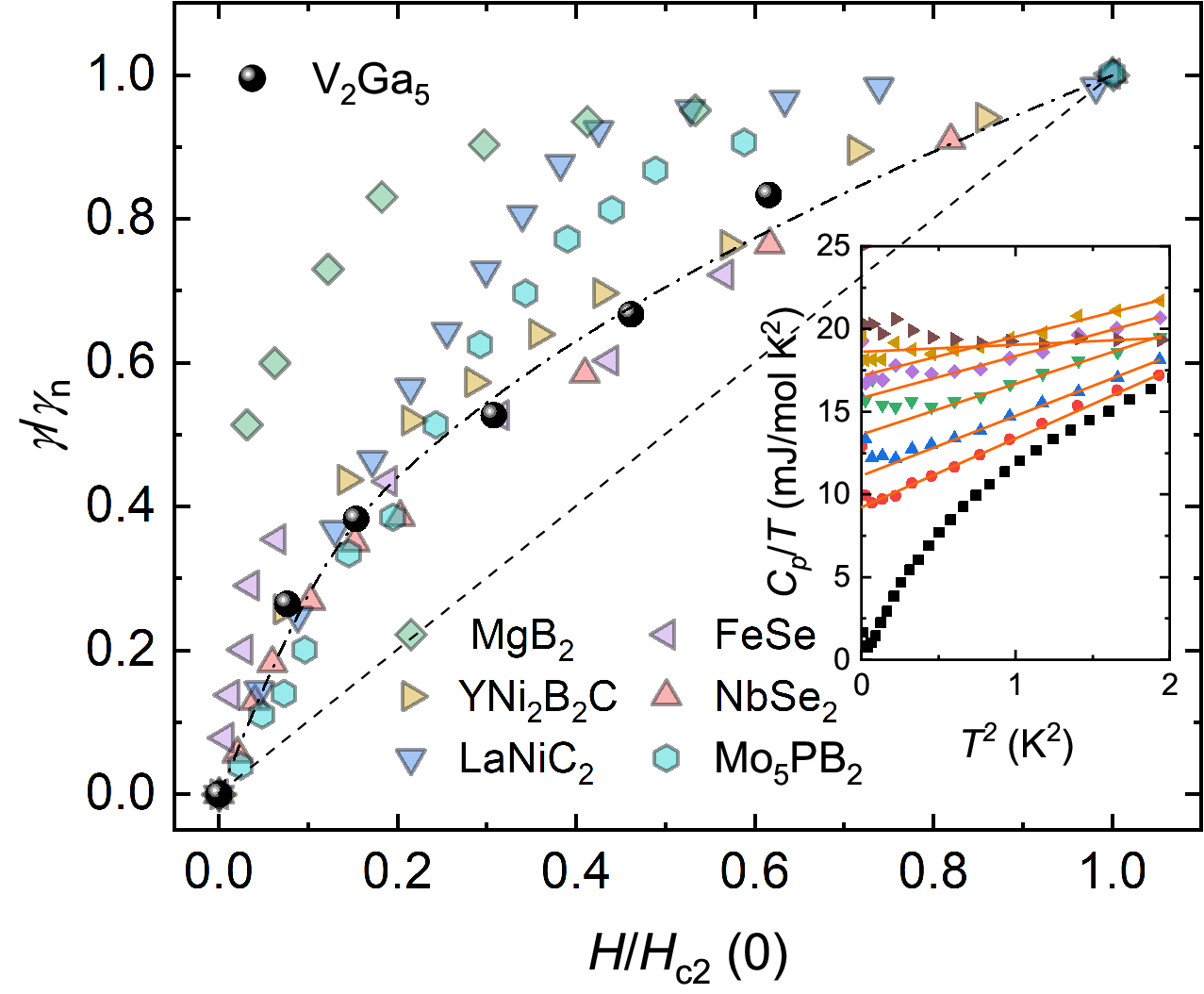}
\caption{Normalized specific heat coefficient $\gamma/\gamma_{n}$ vs. normalized magnetic field $H/H_{c2}(0)$ for V$_{2}$Ga$_{5}$ and other two-gap superconductors. $\gamma$ is obtained from a linear fit of $C/T$ vs. $T^{2}$ and its extrapolation to absolute zero (see inset and text for details). The legends in the inset are identical to the ones of Fig.~\ref{CoverT}(a). Dashed and dashed-dotted lines represent one-gap theoretical models with isotropic and line nodal gap structures, respectively. The data of reference two-gap superconductors are adopted from Refs.~\cite{Bouquet2001,Huang2006,Huang2007,Chen2013,Chen2017,Shang2020}.}  
\label{gammaH} 
\end{figure}
We further studied the vortex excitations in the mixed state by considering the field-dependence of the electronic specific heat coefficient $\gamma(H)$. Figure~\ref{gammaH} shows $\gamma(H)$ of V$_{2}$Ga$_{5}$ where the data points were obtained from the linear extrapolations in a $C/T$ vs. $T^{2}$ plot, as shown in the inset. The fitting of the data was obtained from the relatively high-temperature region to avoid the influence of nuclear contributions from the low-temperature regime. Apparently, $\gamma(H)$ does not follow the linear behavior for an isotropic $s$-wave model. It is however in line with the power-law behavior $\gamma \sim H^{1/2}$ for nodal superconductor, as shown by the dashed-dotted line. Such a power-law dependence of $\gamma(H)$ has also been observed in other two-gap superconductors YNi$_{2}$B$_{2}$C\cite{Huang2006}, NbSe$_{2}$\cite{Huang2007}, and FeSe\cite{Chen2017}. At first glance, the nodal model describing $\gamma(H)$ seems incompatible with the two-gap $s$-wave model that describes $C_{e}/T$ (see Fig.~\ref{CoverT}(b)). However, when the superconducting gap is highly anisotropic, such a discrepancy could be reconciled \cite{Huang2006}. In certain extreme cases among two-gap superconductors, such as MgB$_{2}$ \cite{Bouquet2001} and LaNiC$_{2}$ \cite{Chen2013}, the smaller superconducting gap is closed at the magnetic field strength much smaller than $H_{c2}$. This results a rapid increase of $\gamma$ at low fields and then an gradual saturation with increasing $H$, as shown in Fig.~\ref{gammaH}.

\begin{figure}
\includegraphics[width=\columnwidth]{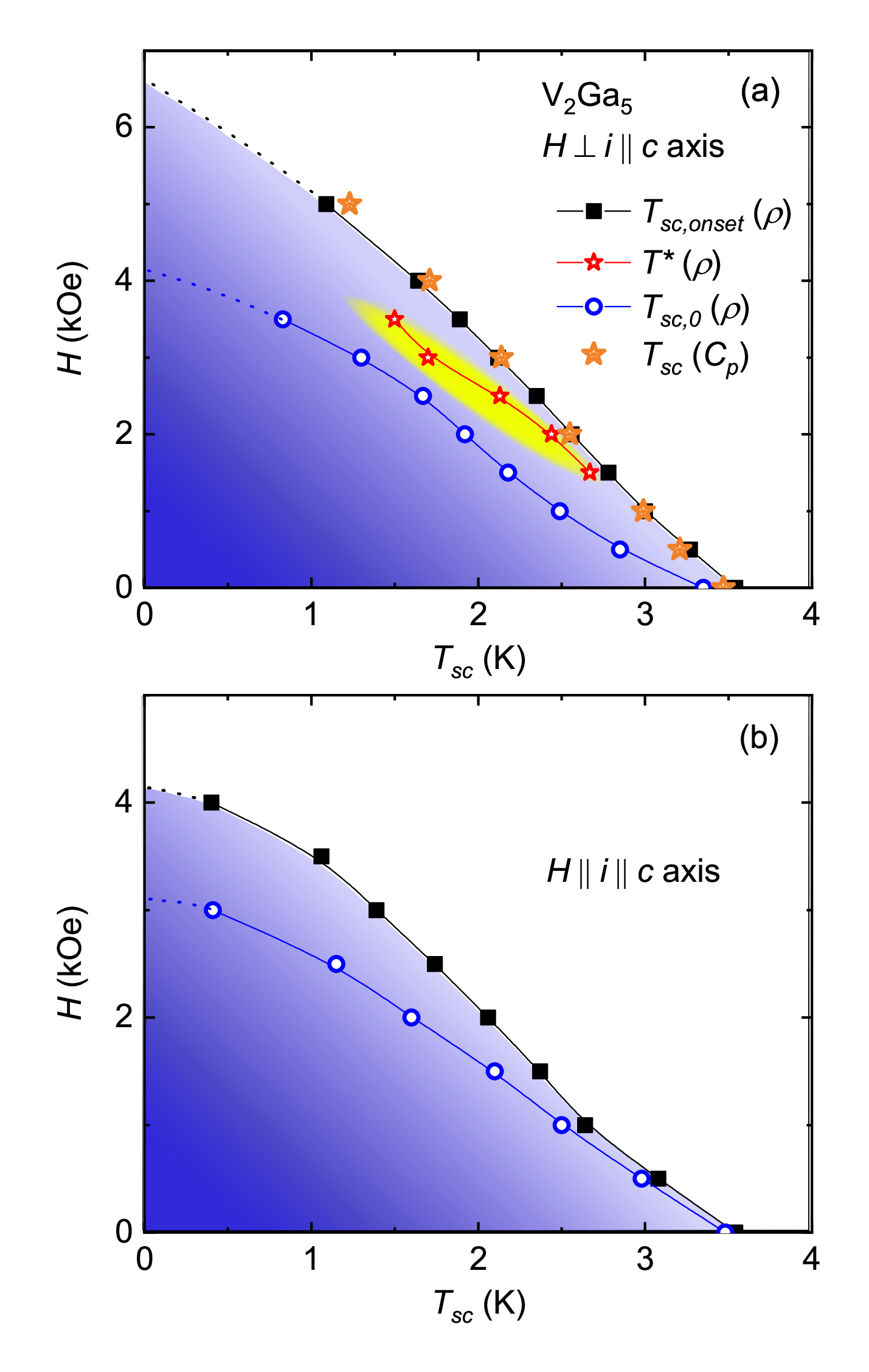}
\caption{(a) $H-T_{sc}$ phase diagram with the magnetic field applied perpendicular to the $c$ axis. A yellow oval represents the region of peak effect. $T^{*}$ denotes the temperature of the resistivity maximum within the superconducting state, i.e., the red arrows in Fig.~\ref{RT}(a). $T_{sc,onset}$ denotes the temperatures where $\rho$ starts to drop. $T_{sc,0}$ denotes the temperature where $\rho$ becomes zero. $T_{sc}$ is the transition temperature determined from specific heat measurements. (b) $H-T_{sc}$ phase diagram with the magnetic field applied parallel to the $c$ axis. Same legends are applied for (a) and (b).}  
\label{phase} 
\end{figure}

We summarize the data from resistivity and specific heat measurements to construct the $H-T_{sc}$ phase diagrams with the magnetic field applied parallel and perpendicular to the $c$-axis, as shown in Fig.~\ref{phase}. The $T_{sc,onset}$ value determined from $\rho(T)$ coincides with the one determined from $C(T)$. 

\begin{figure*}
\includegraphics[width=1.0\linewidth]{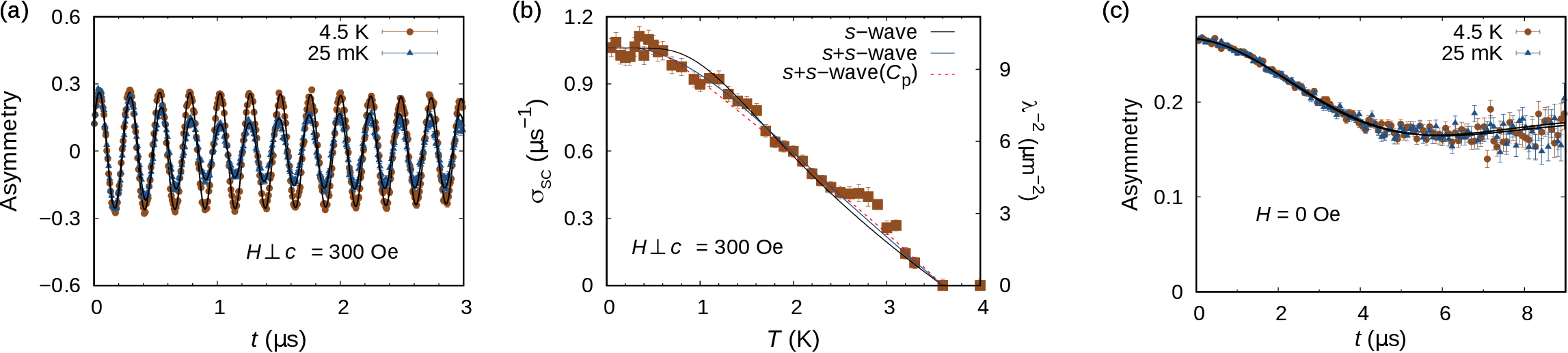}
\caption{(a) Representative transverse field  $\mu$SR spectra for V$_2$Ga$_5$ above and below its $T_{sc}$ collected with a magnetic field applied parallel to the $ab$-plane and muon spin parallel to the $c$-direction. The muon precession in the applied field, which gives rise to the oscillatory behavior in the decay asymmetry, shows an increasing relaxation upon entering the vortex state. The fits to the data (solid lines) are given by Eq.~\ref{TF-fit} as described in the text. These fits yield (b) the superconducting contribution to the muon Gaussian relaxation rate $\sigma_{SC}$ as a function of temperature. An improved fit is obtained by fitting the data with two $s$-wave gaps (blue line) as compared to a single $s$-wave gap (black line). The two $s$-wave gap fit is consistent with the results from specific heat, as shown by the dashed red line. (c) Zero-field $\mu$SR spectra above and below $T_{sc}$ collected with the muon spin parallel to the $ab$-plane show no evidence for time-reversal symmetry breaking. The data are fitted by a Kubo-Toyabe function (solid lines).}  
\label{muSR} 
\end{figure*}
\subsection{Muon spin relaxation}
To further clarify the nature of the superconducting state in V$_2$Ga$_5$, we have performed muon spin relaxation ($\mu$SR) measurements, as summarized in Fig.~\ref{muSR}. Muons are highly sensitive to their local magnetic environment, making them an excellent probe of the vortex state of type II superconductors as well as spontaneous time-reversal symmetry breaking. We begin by investigating the former, where considering the values of $H_{c1,\perp c}$ and $H_{c2,\perp c}$ extracted from our bulk characterization, the sample is expected to be in the vortex state for an applied field of 300~Oe applied $\perp c$ used in our transverse field \mSR\ measurement. In the vortex state, we expect an inhomogenous field distribution due to the presence of a flux line lattice (FLL), which results in a decay of the precession signal as a function of time. The decay in the superconducting state can be clearly seen if we compare the transverse field spectra collected above and below $T_{sc}$, at 25~mK and 4.5~K as shown in Fig.~\ref{muSR}(a).

We fit the transverse field asymmetry spectra with a two term sinusoidal decaying function, 
\begin{equation}
\begin{aligned}
G_{\mathrm{TF}}(t) & =A\left[f \exp \left(\frac{-\sigma^{2} t^{2}}{2}\right) \cos \left(\omega_{1} t+\phi\right)\right. \\
& \left.+(1-f) \exp (-\psi t) \cos \left(\omega_{2} t+\phi\right)\right],
\end{aligned}
\label{TF-fit}
\end{equation}
where the first term captures the signal coming from fraction, $f$, of muons stopping inside the sample, while the second term captures the background signal from the fraction, $1-f$ of muons stopping inside the silver sample holder. The precession frequencies of muons in the sample and the background are given by $\omega_{1}$ and $\omega_{2}$, respectively,  $A$ is the total asymmetry, $\phi$ is the initial phase of the muons, and $\sigma$ and $\psi$ are depolarization rates of the sample and the background signals, respectively. These fits are shown by the solid-lines in Fig.~\ref{muSR}(a).

The superconducting relaxation rate $\left(\sigma_{s c}\right)$ reflects the mean square inhomogeniety in the field experienced by the muons due to the FLL~\cite{brandt1988magnetic}. This quantity was extracted from the fitted Gaussian relaxation rate $\sigma$ by subtracting off a smaller, temperature-independent contribution from randomly oriented nuclear dipole moments $\left(\sigma_{N}\right)$, which give rise to a constant relaxation above $T_{sc}$. The resulting temperature dependence of $\sigma_{sc}$ for V$_2$Ga$_5$ at 300~Oe is plotted in Fig.~\ref{muSR}(b). The penetration depth can be calculated from the relaxation rate~\cite{brandt2003properties}, which then allows a direct calculation of the temperature dependence of the superconducting energy gap~\cite{carrington2003magnetic}. The full protocol for performing these calculations is provided in the Supplemental Material.

In order to further investigate the superconducting gap structure for V$_2$Ga$_5$, we have attempted fits of $\sigma_{SC}$ with both a single $s$-wave gap and two $s$-wave gaps. We note a hump in $\sigma_{SC}$ between 2.5 and 3.0 K that cannot be explained by either of the gap structures under consideration here. We attribute this feature to the movement of vortices and it is likely also related to the peak effect observed in resistivity measurements below $T_{sc}$. In our fits, we therefore exclude this range of temperatures. The resulting single $s$-wave gap and two $s$-wave gaps, as shown with solid black and blue lines in Fig.~\ref{muSR}(b) with the fitting parameters given in the Supplemental Material. The overall quality of the fit is higher for the two gap function, consistent with heat capacity, particularly in the temperature range close to 1 K. For comparison, we show the calculated temperature dependence of $\sigma_{SC}$ based on the two gaps extracted from specific heat by a red dashed-line. Considering the level of noise in the data, we consider the gap values extracted from heat capacity to be more reliable.

Finally, we performed zero field $\mu$SR measurements to look for any sign of spontaneous magnetic fields and time-reversal symmetry breaking as V$_2$Ga$_5$ enters its superconducting state. The magnetic field at the sample position was carefully zeroed following the protocol described in the methods. High statistics zero field spectra were collected above and below $T_{sc}$ with representative spectra shown in Fig.~\ref{muSR}(c). In the absence of any static electronic moments, the muon polarization decay is due to randomly oriented nuclear magnetic moments, as described by the Gaussian Kubo-Toyabe function,
\begin{equation}
G_{\mathrm{KT}}(t)=\frac{1}{3}+\frac{2}{3}\left(1-\sigma^{2} t^{2}\right) e^ {-\frac{\sigma^{2} t^{2}}{2}},
\end{equation}
\noindent
where $\sigma$ reflects the width of the field experienced by muons due to nuclear dipoles. The total fitting function applied to our zero field data is
\begin{equation}
A(t)=A [f G_{\mathrm{KT}}(t) e^{-\Lambda t}+(1+f)],
\end{equation}
\noindent
where $A$ is the sample asymmetry, $f$ represents the asymmetry fraction originating from muons that land in the sample, and $\Lambda$, represents any additional relaxation inside the sample, such as that coming from spontaneous fields inside the superconducting state. The zero field muon spin relaxation rate for V$_2$Ga$_5$ is mainly attributed to a temperature-independent contribution from nuclear dipole moments. No change in $\Lambda$ is observed at temperatures well below $T_{sc}$, as can be evidenced from the comparison of spectra collected at 25~mK and 4.5~K shown in Fig.~\ref{muSR}(c). We conclude from our data that no significant breaking of time-reversal symmetry occurs in the superconducting state. Considering the lack of evidence for TRS breaking and the gap being well fitted with two $s$-wave gaps, we conclude that the superconductivity in V$_2$Ga$_5$ arises from a singlet pairing.

\subsection{Angle-resolved photoemission spectroscopy and band-structure calculations}
\begin{figure*}
\includegraphics[width=0.8\linewidth]{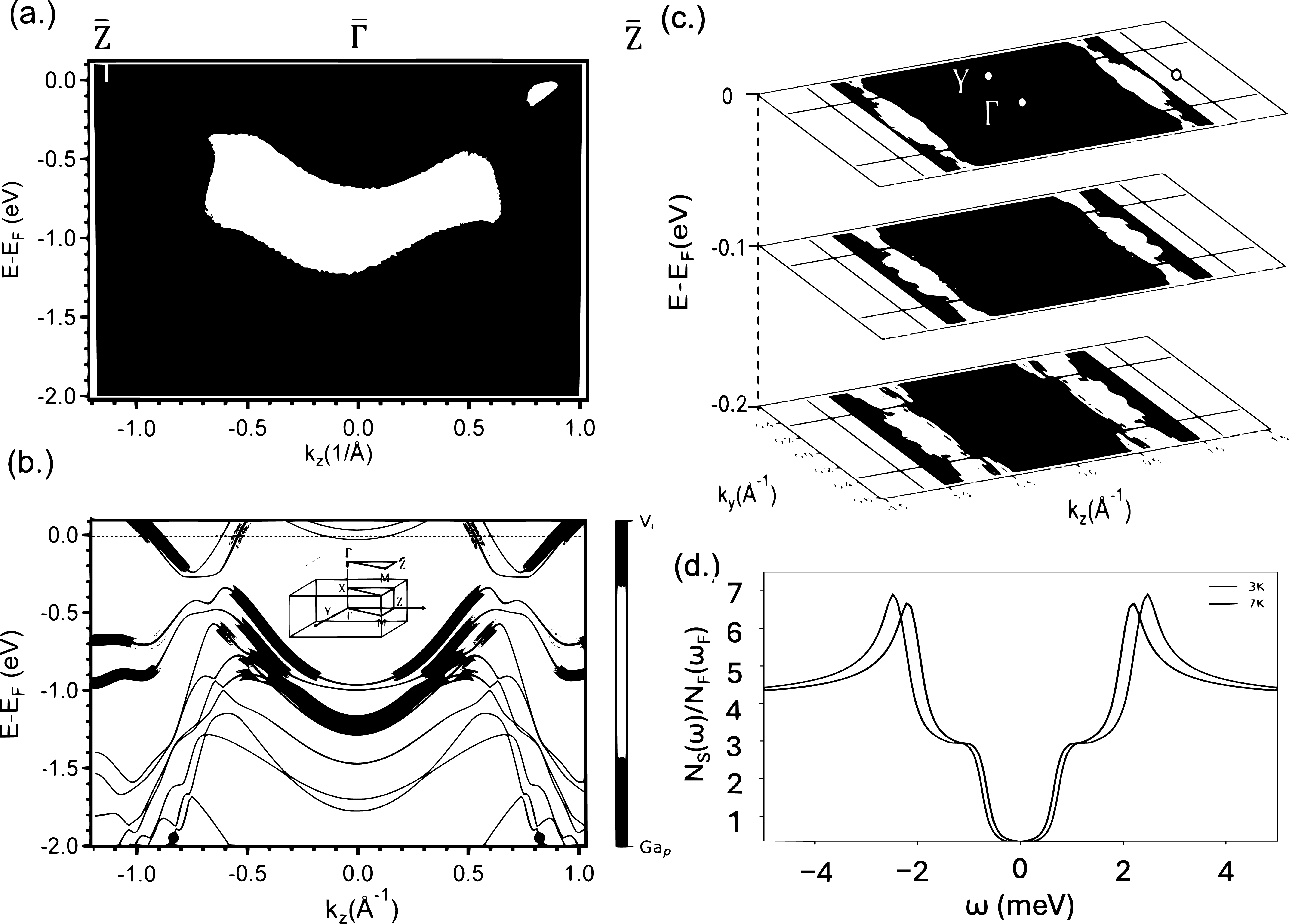}
\caption{(a) ARPES spectrum of V$_2$Ga$_5$ were acquired along the k$_z$ direction at 89 K using a photon energy of 78 eV. (b) Calculated band dispersion of V$_{2}$Ga$_{5}$ along k$_{z}$ direction at (k$_x$, k$_y$) = (0.133 \AA$^{-1}$, 0). The amount of the projected orbital weight of V $d$ orbital and Ga $p$ orbital is represented by red and blue curves, respectively. The color bar indicates the corresponding mixed contribution. (c) Constant energy contours of V$_2$Ga$_5$ at distinct binding energies with respect to the Fermi surface: 0 eV, 0.1 eV, and 0.2 eV, which were measured at 89 K with 78 eV photon energy. (d) Quasiparticle density of states for V$_2$Ga$_5$ in the superconducting state at different temperatures (3~K and 7~K).}
\label{ARPES} 
\end{figure*}

To elucidate the nature of superconductivity in V$_2$Ga$_5$, we conducted ARPES experiments and performed the density functional theory (DFT) calculations to unveil the band structure of V$_2$Ga$_5$. Figure~\ref{ARPES}(a) illustrates the band mapping result of ARPES spectra along the $\Gamma$-Z direction, measured at 89~K utilizing a photon energy of 78 eV. Two discernible bands across the Fermi level were observed: one electron band below $-0.25$ eV of binding energy around the Z point disperses toward the Fermi level, while another, with weaker photoemission intensity, behaves as a parabolic band around the $\Gamma$ point. To gain insight into the out-of-plane band dispersion, we employed a photon energy-dependent ARPES experiment  \cite{Damascelli_2004}, using energies ranging from 52 eV to 86 eV. No significant changes are observed in bands around the Fermi level, suggesting only weakly dispersive behavior along the k$_x$ direction, which is revealed in the Supplemental Material \cite{SM}. Additionally, constant energy mapping in Fig.~\ref{ARPES}(c) demonstrates a less dispersive behavior for these bands around the Fermi level along the k$_y$ direction, without the presence of nodes on the Fermi surface \cite{SM}.

Figure~\ref{ARPES}(b) presents the calculated band structure of V$_2$Ga$_5$ along the $\Gamma$-Z direction at (k$_x$, k$_y$) = (0.133 \AA$^{-1}$, 0), where the orbital contribution for each energy eigenstate is analyzed and the projected weight of V $d$ orbital and Ga $p$ orbital is represented by red and blue curves, respectively. The measured bands in the low binding energy region are mainly composed of the V 3d and Ga 4p orbitals, and the calculated band structure at k$_x$ = 0.133 \AA$^{-1}$ aligns well with the ARPES spectra. As presented in the Supplemental Material \cite{SM}, these bands show less dispersion along the k$_x$ direction, suggesting a quasi-one-dimensional behavior.

Through first-principles calculations, we have identified three bands that cross the Fermi level, and two of them are responsible for the observed two-gap superconductivity. Using the linear response framework (see Supplemental Material \cite{SM}), the electron-phonon coupling strength ($\lambda$) has been calculated. The result yields a theoretical electron-phonon coupling strength of 0.75 and a superconducting critical temperature ($T_{sc}$) of 7.38 K based on the McMillan-Allen-Dynes formula \cite{PhysRev.167.331,PhysRevB.12.905}:
\begin{equation}
T_{sc}=\frac{\omega_{log}}{1.2}\exp\Big[\frac{-1.04(1+\lambda)}{\lambda-\mu_c^*(1+0.62\lambda)}\Big].
\end{equation}
Here, the typical Coulomb pseudopotential value $\mu^*_c$ is set to $0.1$. We have further solved the anisotropic Migdal-Eliashberg equations to investigate the temperature dependence of the superconducting gap. In Fig.~\ref{ARPES}(d), the quasiparticle density of states in the superconducting state is presented at 3~K and 7~K. The quasiparticle density of states clearly shows the two-gap behavior, similar to that in the two-gap superconductor MgB$_{2}$. The calculated $T_{sc}$ and the nodeless Fermi surface are in agreement with our experimental observations, and our theoretical result of the two-gap $s$-wave behavior is also consistent with the recent experimental findings \cite{Xu2024}. However, we note that the theoretical two-gap values are overestimated compared to the experimental ones. This discrepancy requires further theoretical work to reconcile, such as first-principles calculations of the Coulomb pseudopotential.

\section{Summary}
We present a detailed study of the physical properties and electronic structure of the superconductor V$_2$Ga$_5$. From resistivity measurements with the magnetic field applied parallel and perpendicular to the $c$ axis, the $H_{c2}$ vs. $T$ phase diagrams are constructed, where small magnetic anisotropy has been registered. Resistivity under high pressure shows that the superconductivity is suppressed linearly with pressure, suggesting the superconductivity is phonon-mediated. Low-temperature specific heat and $\mu$SR experiments both reveal that the two-gap $s$-wave model best describes the superconducting order parameter. These two gaps can be directly inferred from ARPES measurements and are further supported by band structure calculations. By combining various experiments with theoretical calculations, our findings intricately detail the physical properties and superconductivity of V$_2$Ga$_5$, paving the way for future studies.

\begin{acknowledgements}
We thank R. Pierre and T. Klein for fruitful discussions. We thank Dr. M.-K. Lee and C.-C. Yang at PPMS-16T and SQUID VSM Labs, Instrumentation Center, National Cheng Kung University (NCKU) for technical support. We are grateful to P.-Z. Hsu and L.-J. Chang for the help of Laue diffraction. CLH would like to thank Academia Sinica for "Short-term Domestic Visiting Scholars" program in the Nanomaterial and Low Temperature Physics Laboratory, IOP, Academia Sinica. This work is supported by National Science and Technology Council in Taiwan (grant number NSTC 109-2112-M-006-026-MY3, NSTC 110-2124-M-006-011, and NSTC 112-2112-M-032-010) and the Higher Education Sprout Project, Ministry of Education to the Headquarters of University Advancement at NCKU. WTC acknowledges the National Science and Technology Council in Taiwan, for funding 111-2112-M-002-044-MY3, 112-2124-M-002-012, Academia Sinica project number AS-iMATE-111-12, and the Featured Areas Research Center Program within the framework of the Higher Education Sprout Project by the Ministry of Education of Taiwan 113L9008. LZD would like to thank U.S. Air Force Office of Scientific Research Grants FA9550-15-1-0236 and FA9550-20-1-0068; the T. L. L. Temple Foundation; the John J and Rebecca Moores Endowment; the State of Texas through the Texas Center for Superconductivity at the University of Houston. AMH acknowledges support from the Natural Sciences and Engineering Research Council of Canada (NSERC), the CIFAR Azrieli Global Scholars program, and the Sloan Research Fellowships program. Research at UBC was undertaken, in part, thanks to funding from the Canada First Research Excellence Fund, Quantum Materials and Future Technologies Program.
CCL acknowledges support from the National Center for Theoretical Sciences (NCTS) of Taiwan.
\end{acknowledgements}

%\bibliography{apssamp}

\begin{thebibliography}{43}%
\makeatletter
\providecommand \@ifxundefined [1]{%
 \@ifx{#1\undefined}
}%
\providecommand \@ifnum [1]{%
 \ifnum #1\expandafter \@firstoftwo
 \else \expandafter \@secondoftwo
 \fi
}%
\providecommand \@ifx [1]{%
 \ifx #1\expandafter \@firstoftwo
 \else \expandafter \@secondoftwo
 \fi
}%
\providecommand \natexlab [1]{#1}%
\providecommand \enquote  [1]{``#1''}%
\providecommand \bibnamefont  [1]{#1}%
\providecommand \bibfnamefont [1]{#1}%
\providecommand \citenamefont [1]{#1}%
\providecommand \href@noop [0]{\@secondoftwo}%
\providecommand \href [0]{\begingroup \@sanitize@url \@href}%
\providecommand \@href[1]{\@@startlink{#1}\@@href}%
\providecommand \@@href[1]{\endgroup#1\@@endlink}%
\providecommand \@sanitize@url [0]{\catcode `\\12\catcode `\$12\catcode
  `\&12\catcode `\#12\catcode `\^12\catcode `\_12\catcode `\%12\relax}%
\providecommand \@@startlink[1]{}%
\providecommand \@@endlink[0]{}%
\providecommand \url  [0]{\begingroup\@sanitize@url \@url }%
\providecommand \@url [1]{\endgroup\@href {#1}{\urlprefix }}%
\providecommand \urlprefix  [0]{URL }%
\providecommand \Eprint [0]{\href }%
\providecommand \doibase [0]{http://dx.doi.org/}%
\providecommand \selectlanguage [0]{\@gobble}%
\providecommand \bibinfo  [0]{\@secondoftwo}%
\providecommand \bibfield  [0]{\@secondoftwo}%
\providecommand \translation [1]{[#1]}%
\providecommand \BibitemOpen [0]{}%
\providecommand \bibitemStop [0]{}%
\providecommand \bibitemNoStop [0]{.\EOS\space}%
\providecommand \EOS [0]{\spacefactor3000\relax}%
\providecommand \BibitemShut  [1]{\csname bibitem#1\endcsname}%
\let\auto@bib@innerbib\@empty
%</preamble>
\bibitem [{\citenamefont {Canfield}\ and\ \citenamefont
  {Fisk}(1992)}]{Canfield1992}%
  \BibitemOpen
  \bibfield  {author} {\bibinfo {author} {\bibfnamefont {P.~C.}\ \bibnamefont
  {Canfield}}\ and\ \bibinfo {author} {\bibfnamefont {Z.}~\bibnamefont
  {Fisk}},\ }\bibfield  {title} {\enquote {\bibinfo {title} {Growth of single
  crystals from metallic fluxes},}\ }\href@noop {} {\bibfield  {journal}
  {\bibinfo  {journal} {Philosophical Magazine B}\ }\textbf {\bibinfo {volume}
  {65}},\ \bibinfo {pages} {1117--1123} (\bibinfo {year} {1992})}\BibitemShut
  {NoStop}%
\bibitem [{\citenamefont {Kanatzidis}\ \emph {et~al.}(2005)\citenamefont
  {Kanatzidis}, \citenamefont {Pöttgen},\ and\ \citenamefont
  {Jeitschko}}]{Kanatzidis2005}%
  \BibitemOpen
  \bibfield  {author} {\bibinfo {author} {\bibfnamefont {Mercouri~G.}\
  \bibnamefont {Kanatzidis}}, \bibinfo {author} {\bibfnamefont {Rainer}\
  \bibnamefont {Pöttgen}}, \ and\ \bibinfo {author} {\bibfnamefont {Wolfgang}\
  \bibnamefont {Jeitschko}},\ }\href@noop {} {\bibfield  {journal} {\bibinfo
  {journal} {Angewandte Chemie International Edition}\ }\textbf {\bibinfo
  {volume} {44}},\ \bibinfo {pages} {6996--7023} (\bibinfo {year}
  {2005})}\BibitemShut {NoStop}%
\bibitem [{\citenamefont {Goodman}(1962)}]{Goodman1962}%
  \BibitemOpen
  \bibfield  {author} {\bibinfo {author} {\bibfnamefont {B.B.}\ \bibnamefont
  {Goodman}},\ }\bibfield  {title} {\enquote {\bibinfo {title} {The
  thermodynamic properties of superconducting v3ga},}\ }\href {\doibase
  https://doi.org/10.1016/0031-9163(62)90360-8} {\bibfield  {journal} {\bibinfo
   {journal} {Physics Letters}\ }\textbf {\bibinfo {volume} {1}},\ \bibinfo
  {pages} {215--217} (\bibinfo {year} {1962})}\BibitemShut {NoStop}%
\bibitem [{\citenamefont {Junod}\ \emph {et~al.}(1971)\citenamefont {Junod},
  \citenamefont {Staudenmann}, \citenamefont {Muller},\ and\ \citenamefont
  {Spitzli}}]{Junod1971}%
  \BibitemOpen
  \bibfield  {author} {\bibinfo {author} {\bibfnamefont {A.}~\bibnamefont
  {Junod}}, \bibinfo {author} {\bibfnamefont {J.-L.}\ \bibnamefont
  {Staudenmann}}, \bibinfo {author} {\bibfnamefont {J.}~\bibnamefont {Muller}},
  \ and\ \bibinfo {author} {\bibfnamefont {P.}~\bibnamefont {Spitzli}},\
  }\bibfield  {title} {\enquote {\bibinfo {title} {Superconductivity,
  density-of-states models, and specific heat of a15-type compounds v-ga and
  v-si},}\ }\href@noop {} {\bibfield  {journal} {\bibinfo  {journal} {Journal
  of Low Temperature Physics}\ }\textbf {\bibinfo {volume} {5}},\ \bibinfo
  {pages} {25--43} (\bibinfo {year} {1971})}\BibitemShut {NoStop}%
\bibitem [{\citenamefont {Cruceanu}\ \emph {et~al.}(1974)\citenamefont
  {Cruceanu}, \citenamefont {Antesberger},\ and\ \citenamefont
  {Papastaikoudis}}]{Cruceanu1974}%
  \BibitemOpen
  \bibfield  {author} {\bibinfo {author} {\bibfnamefont {E.}~\bibnamefont
  {Cruceanu}}, \bibinfo {author} {\bibfnamefont {G.}~\bibnamefont
  {Antesberger}}, \ and\ \bibinfo {author} {\bibfnamefont {C.}~\bibnamefont
  {Papastaikoudis}},\ }\href@noop {} {\bibfield  {journal} {\bibinfo  {journal}
  {Solid State Communications}\ }\textbf {\bibinfo {volume} {15}},\ \bibinfo
  {pages} {1047--1049} (\bibinfo {year} {1974})}\BibitemShut {NoStop}%
\bibitem [{\citenamefont {Lobring}\ \emph {et~al.}(2002)\citenamefont
  {Lobring}, \citenamefont {Check}, \citenamefont {Zhang}, \citenamefont {Li},
  \citenamefont {Zheng},\ and\ \citenamefont {Rogacki}}]{Lobring2002}%
  \BibitemOpen
  \bibfield  {author} {\bibinfo {author} {\bibfnamefont {Kim~C}\ \bibnamefont
  {Lobring}}, \bibinfo {author} {\bibfnamefont {Catherine~E}\ \bibnamefont
  {Check}}, \bibinfo {author} {\bibfnamefont {Jianhua}\ \bibnamefont {Zhang}},
  \bibinfo {author} {\bibfnamefont {Shoujian}\ \bibnamefont {Li}}, \bibinfo
  {author} {\bibfnamefont {Chong}\ \bibnamefont {Zheng}}, \ and\ \bibinfo
  {author} {\bibfnamefont {Krzysztof}\ \bibnamefont {Rogacki}},\ }\href@noop {}
  {\bibfield  {journal} {\bibinfo  {journal} {Journal of Alloys and Compounds}\
  }\textbf {\bibinfo {volume} {347}},\ \bibinfo {pages} {72--78} (\bibinfo
  {year} {2002})}\BibitemShut {NoStop}%
\bibitem [{\citenamefont {Teruya}\ \emph {et~al.}(2015)\citenamefont {Teruya},
  \citenamefont {Takeda}, \citenamefont {Nakamura}, \citenamefont {Harima},
  \citenamefont {Haga}, \citenamefont {Uchima}, \citenamefont {Hedo},
  \citenamefont {Nakama},\ and\ \citenamefont {Ōnuki}}]{Teruya2015}%
  \BibitemOpen
  \bibfield  {author} {\bibinfo {author} {\bibfnamefont {Atsushi}\ \bibnamefont
  {Teruya}}, \bibinfo {author} {\bibfnamefont {Masataka}\ \bibnamefont
  {Takeda}}, \bibinfo {author} {\bibfnamefont {Ai}~\bibnamefont {Nakamura}},
  \bibinfo {author} {\bibfnamefont {Hisatomo}\ \bibnamefont {Harima}}, \bibinfo
  {author} {\bibfnamefont {Yoshinori}\ \bibnamefont {Haga}}, \bibinfo {author}
  {\bibfnamefont {Kiyoharu}\ \bibnamefont {Uchima}}, \bibinfo {author}
  {\bibfnamefont {Masato}\ \bibnamefont {Hedo}}, \bibinfo {author}
  {\bibfnamefont {Takao}\ \bibnamefont {Nakama}}, \ and\ \bibinfo {author}
  {\bibfnamefont {Yoshichika}\ \bibnamefont {Ōnuki}},\ }\href@noop {}
  {\bibfield  {journal} {\bibinfo  {journal} {Journal of the Physical Society
  of Japan}\ }\textbf {\bibinfo {volume} {84}},\ \bibinfo {pages} {054703}
  (\bibinfo {year} {2015})}\BibitemShut {NoStop}%
\bibitem [{\citenamefont {Xu}\ \emph {et~al.}(2024)\citenamefont {Xu},
  \citenamefont {Zhao}, \citenamefont {Shen}, \citenamefont {Ratkovski},
  \citenamefont {Ma}, \citenamefont {Zhou}, \citenamefont {Yin}, \citenamefont
  {Li}, \citenamefont {Bangura}, \citenamefont {Cao}, \citenamefont {Wang},
  \citenamefont {Zhu}, \citenamefont {Ke}, \citenamefont {Qian}, \citenamefont
  {Li},\ and\ \citenamefont {Xu}}]{Xu2024}%
  \BibitemOpen
  \bibfield  {author} {\bibinfo {author} {\bibfnamefont {C.~Q.}\ \bibnamefont
  {Xu}}, \bibinfo {author} {\bibfnamefont {C.~C.}\ \bibnamefont {Zhao}},
  \bibinfo {author} {\bibfnamefont {Y.}~\bibnamefont {Shen}}, \bibinfo {author}
  {\bibfnamefont {D.}~\bibnamefont {Ratkovski}}, \bibinfo {author}
  {\bibfnamefont {X.}~\bibnamefont {Ma}}, \bibinfo {author} {\bibfnamefont
  {W.}~\bibnamefont {Zhou}}, \bibinfo {author} {\bibfnamefont {Xunqing}\
  \bibnamefont {Yin}}, \bibinfo {author} {\bibfnamefont {B.}~\bibnamefont
  {Li}}, \bibinfo {author} {\bibfnamefont {A.~F.}\ \bibnamefont {Bangura}},
  \bibinfo {author} {\bibfnamefont {Chao}\ \bibnamefont {Cao}}, \bibinfo
  {author} {\bibfnamefont {Baomin}\ \bibnamefont {Wang}}, \bibinfo {author}
  {\bibfnamefont {Ziming}\ \bibnamefont {Zhu}}, \bibinfo {author}
  {\bibfnamefont {X.}~\bibnamefont {Ke}}, \bibinfo {author} {\bibfnamefont
  {Dong}\ \bibnamefont {Qian}}, \bibinfo {author} {\bibfnamefont {Shiyan}\
  \bibnamefont {Li}}, \ and\ \bibinfo {author} {\bibfnamefont {Xiaofeng}\
  \bibnamefont {Xu}},\ }\bibfield  {title} {\enquote {\bibinfo {title}
  {Multigap nodeless superconductivity in the dirac intermetallic alloy
  ${\mathrm{v}}_{2}{\mathrm{ga}}_{5}$ with one-dimensional vanadium chains},}\
  }\href@noop {} {\bibfield  {journal} {\bibinfo  {journal} {Phys. Rev. B}\
  }\textbf {\bibinfo {volume} {109}},\ \bibinfo {pages} {L100506} (\bibinfo
  {year} {2024})}\BibitemShut {NoStop}%
\bibitem [{\citenamefont {Coelho}(2018)}]{Coelho2018}%
  \BibitemOpen
  \bibfield  {author} {\bibinfo {author} {\bibfnamefont {A.~A.}\ \bibnamefont
  {Coelho}},\ }\bibfield  {title} {\enquote {\bibinfo {title} {Topas and
  topas-academic: an optimization program integrating computer algebra and
  crystallographic objects written in c++},}\ }\href@noop {} {\bibfield
  {journal} {\bibinfo  {journal} {J. Appl. Cryst.}\ }\textbf {\bibinfo {volume}
  {51}},\ \bibinfo {pages} {210--218} (\bibinfo {year} {2018})}\BibitemShut
  {NoStop}%
\bibitem [{SM()}]{SM}%
  \BibitemOpen
  \href@noop {} {\enquote {\bibinfo {title} {See the supplemental material at
  http:xxx for details of $k_x$ dependence of band structure.}}\ }\BibitemShut
  {NoStop}%
\bibitem [{\citenamefont {Morris}\ and\ \citenamefont
  {Heffner}(2003)}]{morris2003method}%
  \BibitemOpen
  \bibfield  {author} {\bibinfo {author} {\bibfnamefont {GD}~\bibnamefont
  {Morris}}\ and\ \bibinfo {author} {\bibfnamefont {RH}~\bibnamefont
  {Heffner}},\ }\bibfield  {title} {\enquote {\bibinfo {title} {A method of
  achieving accurate zero-field conditions using muonium},}\ }\href@noop {}
  {\bibfield  {journal} {\bibinfo  {journal} {Physica B: Condensed Matter}\
  }\textbf {\bibinfo {volume} {326}},\ \bibinfo {pages} {252--254} (\bibinfo
  {year} {2003})}\BibitemShut {NoStop}%
\bibitem [{\citenamefont {Suter}\ and\ \citenamefont
  {Wojek}(2012)}]{suter2012musrfit}%
  \BibitemOpen
  \bibfield  {author} {\bibinfo {author} {\bibfnamefont {A}~\bibnamefont
  {Suter}}\ and\ \bibinfo {author} {\bibfnamefont {BM}~\bibnamefont {Wojek}},\
  }\bibfield  {title} {\enquote {\bibinfo {title} {Musrfit: a free
  platform-independent framework for {$\mu$SR} data analysis},}\ }\href@noop {}
  {\bibfield  {journal} {\bibinfo  {journal} {Physics Procedia}\ }\textbf
  {\bibinfo {volume} {30}},\ \bibinfo {pages} {69--73} (\bibinfo {year}
  {2012})}\BibitemShut {NoStop}%
\bibitem [{\citenamefont {Giannozzi}\ \emph {et~al.}(2017)\citenamefont
  {Giannozzi}, \citenamefont {Andreussi}, \citenamefont {Brumme}, \citenamefont
  {Bunau}, \citenamefont {Nardelli}, \citenamefont {Calandra}, \citenamefont
  {Car}, \citenamefont {Cavazzoni}, \citenamefont {Ceresoli}, \citenamefont
  {Cococcioni}, \citenamefont {Colonna}, \citenamefont {Carnimeo},
  \citenamefont {Corso}, \citenamefont {de~Gironcoli}, \citenamefont {Delugas},
  \citenamefont {DiStasio}, \citenamefont {Ferretti}, \citenamefont {Floris},
  \citenamefont {Fratesi}, \citenamefont {Fugallo}, \citenamefont {Gebauer},
  \citenamefont {Gerstmann}, \citenamefont {Giustino}, \citenamefont {Gorni},
  \citenamefont {Jia}, \citenamefont {Kawamura}, \citenamefont {Ko},
  \citenamefont {Kokalj}, \citenamefont {Küçükbenli}, \citenamefont
  {Lazzeri}, \citenamefont {Marsili}, \citenamefont {Marzari}, \citenamefont
  {Mauri}, \citenamefont {Nguyen}, \citenamefont {Nguyen}, \citenamefont {de-la
  Roza}, \citenamefont {Paulatto}, \citenamefont {Poncé}, \citenamefont
  {Rocca}, \citenamefont {Sabatini}, \citenamefont {Santra}, \citenamefont
  {Schlipf}, \citenamefont {Seitsonen}, \citenamefont {Smogunov}, \citenamefont
  {Timrov}, \citenamefont {Thonhauser}, \citenamefont {Umari}, \citenamefont
  {Vast}, \citenamefont {Wu},\ and\ \citenamefont {Baroni}}]{Giannozzi_2017}%
  \BibitemOpen
  \bibfield  {author} {\bibinfo {author} {\bibfnamefont {P}~\bibnamefont
  {Giannozzi}}, \bibinfo {author} {\bibfnamefont {O}~\bibnamefont {Andreussi}},
  \bibinfo {author} {\bibfnamefont {T}~\bibnamefont {Brumme}}, \bibinfo
  {author} {\bibfnamefont {O}~\bibnamefont {Bunau}}, \bibinfo {author}
  {\bibfnamefont {M~Buongiorno}\ \bibnamefont {Nardelli}}, \bibinfo {author}
  {\bibfnamefont {M}~\bibnamefont {Calandra}}, \bibinfo {author} {\bibfnamefont
  {R}~\bibnamefont {Car}}, \bibinfo {author} {\bibfnamefont {C}~\bibnamefont
  {Cavazzoni}}, \bibinfo {author} {\bibfnamefont {D}~\bibnamefont {Ceresoli}},
  \bibinfo {author} {\bibfnamefont {M}~\bibnamefont {Cococcioni}}, \bibinfo
  {author} {\bibfnamefont {N}~\bibnamefont {Colonna}}, \bibinfo {author}
  {\bibfnamefont {I}~\bibnamefont {Carnimeo}}, \bibinfo {author} {\bibfnamefont
  {A~Dal}\ \bibnamefont {Corso}}, \bibinfo {author} {\bibfnamefont
  {S}~\bibnamefont {de~Gironcoli}}, \bibinfo {author} {\bibfnamefont
  {P}~\bibnamefont {Delugas}}, \bibinfo {author} {\bibfnamefont {R~A}\
  \bibnamefont {DiStasio}}, \bibinfo {author} {\bibfnamefont {A}~\bibnamefont
  {Ferretti}}, \bibinfo {author} {\bibfnamefont {A}~\bibnamefont {Floris}},
  \bibinfo {author} {\bibfnamefont {G}~\bibnamefont {Fratesi}}, \bibinfo
  {author} {\bibfnamefont {G}~\bibnamefont {Fugallo}}, \bibinfo {author}
  {\bibfnamefont {R}~\bibnamefont {Gebauer}}, \bibinfo {author} {\bibfnamefont
  {U}~\bibnamefont {Gerstmann}}, \bibinfo {author} {\bibfnamefont
  {F}~\bibnamefont {Giustino}}, \bibinfo {author} {\bibfnamefont
  {T}~\bibnamefont {Gorni}}, \bibinfo {author} {\bibfnamefont {J}~\bibnamefont
  {Jia}}, \bibinfo {author} {\bibfnamefont {M}~\bibnamefont {Kawamura}},
  \bibinfo {author} {\bibfnamefont {H-Y}\ \bibnamefont {Ko}}, \bibinfo {author}
  {\bibfnamefont {A}~\bibnamefont {Kokalj}}, \bibinfo {author} {\bibfnamefont
  {E}~\bibnamefont {Küçükbenli}}, \bibinfo {author} {\bibfnamefont
  {M}~\bibnamefont {Lazzeri}}, \bibinfo {author} {\bibfnamefont
  {M}~\bibnamefont {Marsili}}, \bibinfo {author} {\bibfnamefont
  {N}~\bibnamefont {Marzari}}, \bibinfo {author} {\bibfnamefont
  {F}~\bibnamefont {Mauri}}, \bibinfo {author} {\bibfnamefont {N~L}\
  \bibnamefont {Nguyen}}, \bibinfo {author} {\bibfnamefont {H-V}\ \bibnamefont
  {Nguyen}}, \bibinfo {author} {\bibfnamefont {A~Otero}\ \bibnamefont {de-la
  Roza}}, \bibinfo {author} {\bibfnamefont {L}~\bibnamefont {Paulatto}},
  \bibinfo {author} {\bibfnamefont {S}~\bibnamefont {Poncé}}, \bibinfo
  {author} {\bibfnamefont {D}~\bibnamefont {Rocca}}, \bibinfo {author}
  {\bibfnamefont {R}~\bibnamefont {Sabatini}}, \bibinfo {author} {\bibfnamefont
  {B}~\bibnamefont {Santra}}, \bibinfo {author} {\bibfnamefont {M}~\bibnamefont
  {Schlipf}}, \bibinfo {author} {\bibfnamefont {A~P}\ \bibnamefont
  {Seitsonen}}, \bibinfo {author} {\bibfnamefont {A}~\bibnamefont {Smogunov}},
  \bibinfo {author} {\bibfnamefont {I}~\bibnamefont {Timrov}}, \bibinfo
  {author} {\bibfnamefont {T}~\bibnamefont {Thonhauser}}, \bibinfo {author}
  {\bibfnamefont {P}~\bibnamefont {Umari}}, \bibinfo {author} {\bibfnamefont
  {N}~\bibnamefont {Vast}}, \bibinfo {author} {\bibfnamefont {X}~\bibnamefont
  {Wu}}, \ and\ \bibinfo {author} {\bibfnamefont {S}~\bibnamefont {Baroni}},\
  }\bibfield  {title} {\enquote {\bibinfo {title} {Advanced capabilities for
  materials modelling with quantum espresso},}\ }\href {\doibase
  10.1088/1361-648X/aa8f79} {\bibfield  {journal} {\bibinfo  {journal} {Journal
  of Physics: Condensed Matter}\ }\textbf {\bibinfo {volume} {29}},\ \bibinfo
  {pages} {465901} (\bibinfo {year} {2017})}\BibitemShut {NoStop}%
\bibitem [{\citenamefont {Ponc\'e}\ \emph {et~al.}(2016)\citenamefont
  {Ponc\'e}, \citenamefont {Margine}, \citenamefont {Verdi},\ and\
  \citenamefont {Giustino}}]{Ponce2016}%
  \BibitemOpen
  \bibfield  {author} {\bibinfo {author} {\bibfnamefont {S.}~\bibnamefont
  {Ponc\'e}}, \bibinfo {author} {\bibfnamefont {E.R.}\ \bibnamefont {Margine}},
  \bibinfo {author} {\bibfnamefont {C.}~\bibnamefont {Verdi}}, \ and\ \bibinfo
  {author} {\bibfnamefont {F.}~\bibnamefont {Giustino}},\ }\bibfield  {title}
  {\enquote {\bibinfo {title} {Epw: Electron–phonon coupling, transport and
  superconducting properties using maximally localized wannier functions},}\
  }\href {\doibase https://doi.org/10.1016/j.cpc.2016.07.028} {\bibfield
  {journal} {\bibinfo  {journal} {Computer Physics Communications}\ }\textbf
  {\bibinfo {volume} {209}},\ \bibinfo {pages} {116 -- 133} (\bibinfo {year}
  {2016})}\BibitemShut {NoStop}%
\bibitem [{\citenamefont {Vucht}\ \emph {et~al.}(1964)\citenamefont {Vucht},
  \citenamefont {Bruning}, \citenamefont {Donkersloot},\ and\ \citenamefont
  {Gomes~de Mesquita}}]{Vucht1964}%
  \BibitemOpen
  \bibfield  {author} {\bibinfo {author} {\bibfnamefont {J.~H. N.~van}\
  \bibnamefont {Vucht}}, \bibinfo {author} {\bibfnamefont {H.~A. C.~M.}\
  \bibnamefont {Bruning}}, \bibinfo {author} {\bibfnamefont {H.~C.}\
  \bibnamefont {Donkersloot}}, \ and\ \bibinfo {author} {\bibfnamefont {A.~H.}\
  \bibnamefont {Gomes~de Mesquita}},\ }\bibfield  {title} {\enquote {\bibinfo
  {title} {The system vanadium-gallium},}\ }\href@noop {} {\bibfield  {journal}
  {\bibinfo  {journal} {Philips Research Reports}\ }\textbf {\bibinfo {volume}
  {19}},\ \bibinfo {pages} {407--421} (\bibinfo {year} {1964})}\BibitemShut
  {NoStop}%
\bibitem [{\citenamefont {Cullity}\ and\ \citenamefont
  {Graham}(2008)}]{Cullity2008}%
  \BibitemOpen
  \bibfield  {author} {\bibinfo {author} {\bibfnamefont {B.~D.}\ \bibnamefont
  {Cullity}}\ and\ \bibinfo {author} {\bibfnamefont {C.~D.}\ \bibnamefont
  {Graham}},\ }in\ \href@noop {} {\emph {\bibinfo {booktitle} {Introduction to
  Magnetic Materials}}}\ (\bibinfo  {publisher} {John Wiley \& Sons},\ \bibinfo
  {year} {2008})\ Chap.~\bibinfo {chapter} {16}, pp.\ \bibinfo {pages}
  {517--526}\BibitemShut {NoStop}%
\bibitem [{\citenamefont {Kittel}(2004)}]{Kittel}%
  \BibitemOpen
  \bibfield  {author} {\bibinfo {author} {\bibfnamefont {C.}~\bibnamefont
  {Kittel}},\ }\href@noop {} {\emph {\bibinfo {title} {Introduction to solid
  state physics}}}\ (\bibinfo  {publisher} {NY: John Wiley \& Sons},\ \bibinfo
  {year} {2004})\BibitemShut {NoStop}%
\bibitem [{\citenamefont {Tomy}\ \emph {et~al.}(1997)\citenamefont {Tomy},
  \citenamefont {Balakrishnan},\ and\ \citenamefont {Paul}}]{Tomy1997}%
  \BibitemOpen
  \bibfield  {author} {\bibinfo {author} {\bibfnamefont {C.~V.}\ \bibnamefont
  {Tomy}}, \bibinfo {author} {\bibfnamefont {G.}~\bibnamefont {Balakrishnan}},
  \ and\ \bibinfo {author} {\bibfnamefont {D.~McK.}\ \bibnamefont {Paul}},\
  }\href@noop {} {\bibfield  {journal} {\bibinfo  {journal} {Phys. Rev. B}\
  }\textbf {\bibinfo {volume} {56}},\ \bibinfo {pages} {8346--8350} (\bibinfo
  {year} {1997})}\BibitemShut {NoStop}%
\bibitem [{\citenamefont {Autler}\ \emph {et~al.}(1962)\citenamefont {Autler},
  \citenamefont {Rosenblum},\ and\ \citenamefont {Gooen}}]{Autler1962}%
  \BibitemOpen
  \bibfield  {author} {\bibinfo {author} {\bibfnamefont {S.~H.}\ \bibnamefont
  {Autler}}, \bibinfo {author} {\bibfnamefont {E.~S.}\ \bibnamefont
  {Rosenblum}}, \ and\ \bibinfo {author} {\bibfnamefont {K.~H.}\ \bibnamefont
  {Gooen}},\ }\href@noop {} {\bibfield  {journal} {\bibinfo  {journal} {Phys.
  Rev. Lett.}\ }\textbf {\bibinfo {volume} {9}},\ \bibinfo {pages} {489--493}
  (\bibinfo {year} {1962})}\BibitemShut {NoStop}%
\bibitem [{\citenamefont {Higgins}\ and\ \citenamefont
  {Bhattacharya}(1996)}]{Higgins1996}%
  \BibitemOpen
  \bibfield  {author} {\bibinfo {author} {\bibfnamefont {Mark~J.}\ \bibnamefont
  {Higgins}}\ and\ \bibinfo {author} {\bibfnamefont {S.}~\bibnamefont
  {Bhattacharya}},\ }\href@noop {} {\bibfield  {journal} {\bibinfo  {journal}
  {Physica C: Superconductivity and its Applications}\ }\textbf {\bibinfo
  {volume} {257}},\ \bibinfo {pages} {232--254} (\bibinfo {year}
  {1996})}\BibitemShut {NoStop}%
\bibitem [{\citenamefont {Hedo}\ \emph {et~al.}(1998)\citenamefont {Hedo},
  \citenamefont {Kobayashi}, \citenamefont {Inada}, \citenamefont {Yamamoto},
  \citenamefont {Haga}, \citenamefont {Suzuki}, \citenamefont {Metoki},
  \citenamefont {Ōnuki}, \citenamefont {Sugawara}, \citenamefont {Sato},
  \citenamefont {Tenya}, \citenamefont {Tayama}, \citenamefont {Amitsuka},\
  and\ \citenamefont {Sakakibara}}]{Hedo1998}%
  \BibitemOpen
  \bibfield  {author} {\bibinfo {author} {\bibfnamefont {Masato}\ \bibnamefont
  {Hedo}}, \bibinfo {author} {\bibfnamefont {Yoshihiko}\ \bibnamefont
  {Kobayashi}}, \bibinfo {author} {\bibfnamefont {Yoshihiko}\ \bibnamefont
  {Inada}}, \bibinfo {author} {\bibfnamefont {Etsuji}\ \bibnamefont
  {Yamamoto}}, \bibinfo {author} {\bibfnamefont {Yoshinori}\ \bibnamefont
  {Haga}}, \bibinfo {author} {\bibfnamefont {Jun-ichi}\ \bibnamefont {Suzuki}},
  \bibinfo {author} {\bibfnamefont {Naoto}\ \bibnamefont {Metoki}}, \bibinfo
  {author} {\bibfnamefont {Yoshichika}\ \bibnamefont {Ōnuki}}, \bibinfo
  {author} {\bibfnamefont {Hitoshi}\ \bibnamefont {Sugawara}}, \bibinfo
  {author} {\bibfnamefont {Hideyuki}\ \bibnamefont {Sato}}, \bibinfo {author}
  {\bibfnamefont {Kenichi}\ \bibnamefont {Tenya}}, \bibinfo {author}
  {\bibfnamefont {Takashi}\ \bibnamefont {Tayama}}, \bibinfo {author}
  {\bibfnamefont {Hiroshi}\ \bibnamefont {Amitsuka}}, \ and\ \bibinfo {author}
  {\bibfnamefont {Toshiro}\ \bibnamefont {Sakakibara}},\ }\href@noop {}
  {\bibfield  {journal} {\bibinfo  {journal} {Journal of the Physical Society
  of Japan}\ }\textbf {\bibinfo {volume} {67}},\ \bibinfo {pages} {3561--3569}
  (\bibinfo {year} {1998})}\BibitemShut {NoStop}%
\bibitem [{\citenamefont {Chaudhary}\ \emph {et~al.}(2001)\citenamefont
  {Chaudhary}, \citenamefont {Rajarajan}, \citenamefont {Singh}, \citenamefont
  {Roy},\ and\ \citenamefont {Chaddah}}]{Chaudhary2001}%
  \BibitemOpen
  \bibfield  {author} {\bibinfo {author} {\bibfnamefont {S.}~\bibnamefont
  {Chaudhary}}, \bibinfo {author} {\bibfnamefont {A.K.}\ \bibnamefont
  {Rajarajan}}, \bibinfo {author} {\bibfnamefont {K.J.}\ \bibnamefont {Singh}},
  \bibinfo {author} {\bibfnamefont {S.B.}\ \bibnamefont {Roy}}, \ and\ \bibinfo
  {author} {\bibfnamefont {P.}~\bibnamefont {Chaddah}},\ }\href@noop {}
  {\bibfield  {journal} {\bibinfo  {journal} {Physica C: Superconductivity}\
  }\textbf {\bibinfo {volume} {353}},\ \bibinfo {pages} {29--37} (\bibinfo
  {year} {2001})}\BibitemShut {NoStop}%
\bibitem [{\citenamefont {Jang}\ \emph {et~al.}(2008)\citenamefont {Jang},
  \citenamefont {Kang}, \citenamefont {Lee}, \citenamefont {Lee}, \citenamefont
  {Jo}, \citenamefont {Jung}, \citenamefont {Cho},\ and\ \citenamefont
  {Lee}}]{Lee2008}%
  \BibitemOpen
  \bibfield  {author} {\bibinfo {author} {\bibfnamefont {H-S Lee D-J}\
  \bibnamefont {Jang}}, \bibinfo {author} {\bibfnamefont {B}~\bibnamefont
  {Kang}}, \bibinfo {author} {\bibfnamefont {H-G}\ \bibnamefont {Lee}},
  \bibinfo {author} {\bibfnamefont {I~J}\ \bibnamefont {Lee}}, \bibinfo
  {author} {\bibfnamefont {Y}~\bibnamefont {Jo}}, \bibinfo {author}
  {\bibfnamefont {M-H}\ \bibnamefont {Jung}}, \bibinfo {author} {\bibfnamefont
  {M-H}\ \bibnamefont {Cho}}, \ and\ \bibinfo {author} {\bibfnamefont {S-I}\
  \bibnamefont {Lee}},\ }\href@noop {} {\bibfield  {journal} {\bibinfo
  {journal} {New Journal of Physics}\ }\textbf {\bibinfo {volume} {10}},\
  \bibinfo {pages} {063003} (\bibinfo {year} {2008})}\BibitemShut {NoStop}%
\bibitem [{\citenamefont {Kaluarachchi}\ \emph {et~al.}(2016)\citenamefont
  {Kaluarachchi}, \citenamefont {Lin}, \citenamefont {Xie}, \citenamefont
  {Taufour}, \citenamefont {Bud'ko}, \citenamefont {Miller},\ and\
  \citenamefont {Canfield}}]{Kaluarachchi2016}%
  \BibitemOpen
  \bibfield  {author} {\bibinfo {author} {\bibfnamefont {Udhara~S.}\
  \bibnamefont {Kaluarachchi}}, \bibinfo {author} {\bibfnamefont {Qisheng}\
  \bibnamefont {Lin}}, \bibinfo {author} {\bibfnamefont {Weiwei}\ \bibnamefont
  {Xie}}, \bibinfo {author} {\bibfnamefont {Valentin}\ \bibnamefont {Taufour}},
  \bibinfo {author} {\bibfnamefont {Sergey~L.}\ \bibnamefont {Bud'ko}},
  \bibinfo {author} {\bibfnamefont {Gordon~J.}\ \bibnamefont {Miller}}, \ and\
  \bibinfo {author} {\bibfnamefont {Paul~C.}\ \bibnamefont {Canfield}},\
  }\href@noop {} {\bibfield  {journal} {\bibinfo  {journal} {Phys. Rev. B}\
  }\textbf {\bibinfo {volume} {93}},\ \bibinfo {pages} {094524} (\bibinfo
  {year} {2016})}\BibitemShut {NoStop}%
\bibitem [{\citenamefont {Willa}\ \emph {et~al.}(2018)\citenamefont {Willa},
  \citenamefont {Koshelev}, \citenamefont {Sadovskyy},\ and\ \citenamefont
  {Glatz}}]{Willa2018}%
  \BibitemOpen
  \bibfield  {author} {\bibinfo {author} {\bibfnamefont {Roland}\ \bibnamefont
  {Willa}}, \bibinfo {author} {\bibfnamefont {Alexei~E.}\ \bibnamefont
  {Koshelev}}, \bibinfo {author} {\bibfnamefont {Ivan~A.}\ \bibnamefont
  {Sadovskyy}}, \ and\ \bibinfo {author} {\bibfnamefont {Andreas}\ \bibnamefont
  {Glatz}},\ }\bibfield  {title} {\enquote {\bibinfo {title} {Peak effect due
  to competing vortex ground states in superconductors with large
  inclusions},}\ }\href {\doibase 10.1103/PhysRevB.98.054517} {\bibfield
  {journal} {\bibinfo  {journal} {Phys. Rev. B}\ }\textbf {\bibinfo {volume}
  {98}},\ \bibinfo {pages} {054517} (\bibinfo {year} {2018})}\BibitemShut
  {NoStop}%
\bibitem [{\citenamefont {Lorenz}\ and\ \citenamefont
  {Chu}(2005)}]{Lorenz2005}%
  \BibitemOpen
  \bibfield  {author} {\bibinfo {author} {\bibfnamefont {B.}~\bibnamefont
  {Lorenz}}\ and\ \bibinfo {author} {\bibfnamefont {C.~W.}\ \bibnamefont
  {Chu}},\ }in\ \href@noop {} {\emph {\bibinfo {booktitle} {High Pressure
  Effects on Superconductivity}}}\ (\bibinfo  {publisher} {Frontiers in
  Superconducting Materials. Springer, Berlin, Heidelberg},\ \bibinfo {year}
  {2005})\BibitemShut {NoStop}%
\bibitem [{\citenamefont {Larson}\ and\ \citenamefont
  {Dreele}(1999)}]{Grimvall1999}%
  \BibitemOpen
  \bibfield  {author} {\bibinfo {author} {\bibfnamefont {A.~C.}\ \bibnamefont
  {Larson}}\ and\ \bibinfo {author} {\bibfnamefont {R.~B.~V.}\ \bibnamefont
  {Dreele}},\ }in\ \href@noop {} {\emph {\bibinfo {booktitle} {Thermophysical
  Properties of Materials}}}\ (\bibinfo  {publisher} {Los Alamos National
  Laboratory, Los Alamos, NM},\ \bibinfo {year} {1999})\ p.\ \bibinfo {pages}
  {p. 324ff}\BibitemShut {NoStop}%
\bibitem [{\citenamefont {Man}\ and\ \citenamefont {Huang}(2011)}]{Man2011}%
  \BibitemOpen
  \bibfield  {author} {\bibinfo {author} {\bibfnamefont {Chi-Sing}\
  \bibnamefont {Man}}\ and\ \bibinfo {author} {\bibfnamefont {Mojia}\
  \bibnamefont {Huang}},\ }\href@noop {} {\bibfield  {journal} {\bibinfo
  {journal} {Journal of Elasticity}\ }\textbf {\bibinfo {volume} {105}},\
  \bibinfo {pages} {29--48} (\bibinfo {year} {2011})}\BibitemShut {NoStop}%
\bibitem [{\citenamefont {Grube}()}]{Grube2022}%
  \BibitemOpen
  \bibfield  {author} {\bibinfo {author} {\bibfnamefont {K.}~\bibnamefont
  {Grube}},\ }\href@noop {} {}\bibinfo {howpublished} {private
  communication}\BibitemShut {NoStop}%
\bibitem [{\citenamefont {Rai}\ \emph {et~al.}(2018)\citenamefont {Rai},
  \citenamefont {Chikara}, \citenamefont {Ding}, \citenamefont {Oswald},
  \citenamefont {Sch\"onemann}, \citenamefont {Loganathan}, \citenamefont
  {Hallas}, \citenamefont {Cao}, \citenamefont {Stavinoha}, \citenamefont
  {Chen}, \citenamefont {Man}, \citenamefont {Carr}, \citenamefont {Singleton},
  \citenamefont {Zapf}, \citenamefont {Benavides}, \citenamefont {Chan},
  \citenamefont {Zhang}, \citenamefont {Rhodes}, \citenamefont {Chiu},
  \citenamefont {Balicas}, \citenamefont {Aczel}, \citenamefont {Huang},
  \citenamefont {Lynn}, \citenamefont {Gaudet}, \citenamefont {Sokolov},
  \citenamefont {Walker}, \citenamefont {Adroja}, \citenamefont {Dai},
  \citenamefont {Nevidomskyy}, \citenamefont {Huang},\ and\ \citenamefont
  {Morosan}}]{Rai2018}%
  \BibitemOpen
  \bibfield  {author} {\bibinfo {author} {\bibfnamefont {Binod~K.}\
  \bibnamefont {Rai}}, \bibinfo {author} {\bibfnamefont {S.}~\bibnamefont
  {Chikara}}, \bibinfo {author} {\bibfnamefont {Xiaxin}\ \bibnamefont {Ding}},
  \bibinfo {author} {\bibfnamefont {Iain W.~H.}\ \bibnamefont {Oswald}},
  \bibinfo {author} {\bibfnamefont {R.}~\bibnamefont {Sch\"onemann}}, \bibinfo
  {author} {\bibfnamefont {V.}~\bibnamefont {Loganathan}}, \bibinfo {author}
  {\bibfnamefont {A.~M.}\ \bibnamefont {Hallas}}, \bibinfo {author}
  {\bibfnamefont {H.~B.}\ \bibnamefont {Cao}}, \bibinfo {author} {\bibfnamefont
  {Macy}\ \bibnamefont {Stavinoha}}, \bibinfo {author} {\bibfnamefont
  {T.}~\bibnamefont {Chen}}, \bibinfo {author} {\bibfnamefont {Haoran}\
  \bibnamefont {Man}}, \bibinfo {author} {\bibfnamefont {Scott}\ \bibnamefont
  {Carr}}, \bibinfo {author} {\bibfnamefont {John}\ \bibnamefont {Singleton}},
  \bibinfo {author} {\bibfnamefont {Vivien}\ \bibnamefont {Zapf}}, \bibinfo
  {author} {\bibfnamefont {Katherine~A.}\ \bibnamefont {Benavides}}, \bibinfo
  {author} {\bibfnamefont {Julia~Y.}\ \bibnamefont {Chan}}, \bibinfo {author}
  {\bibfnamefont {Q.~R.}\ \bibnamefont {Zhang}}, \bibinfo {author}
  {\bibfnamefont {D.}~\bibnamefont {Rhodes}}, \bibinfo {author} {\bibfnamefont
  {Y.~C.}\ \bibnamefont {Chiu}}, \bibinfo {author} {\bibfnamefont {Luis}\
  \bibnamefont {Balicas}}, \bibinfo {author} {\bibfnamefont {A.~A.}\
  \bibnamefont {Aczel}}, \bibinfo {author} {\bibfnamefont {Q.}~\bibnamefont
  {Huang}}, \bibinfo {author} {\bibfnamefont {Jeffrey~W.}\ \bibnamefont
  {Lynn}}, \bibinfo {author} {\bibfnamefont {J.}~\bibnamefont {Gaudet}},
  \bibinfo {author} {\bibfnamefont {D.~A.}\ \bibnamefont {Sokolov}}, \bibinfo
  {author} {\bibfnamefont {H.~C.}\ \bibnamefont {Walker}}, \bibinfo {author}
  {\bibfnamefont {D.~T.}\ \bibnamefont {Adroja}}, \bibinfo {author}
  {\bibfnamefont {Pengcheng}\ \bibnamefont {Dai}}, \bibinfo {author}
  {\bibfnamefont {Andriy~H.}\ \bibnamefont {Nevidomskyy}}, \bibinfo {author}
  {\bibfnamefont {C.-L.}\ \bibnamefont {Huang}}, \ and\ \bibinfo {author}
  {\bibfnamefont {E.}~\bibnamefont {Morosan}},\ }\bibfield  {title} {\enquote
  {\bibinfo {title} {Anomalous metamagnetism in the low carrier density kondo
  lattice ${\mathrm{ybrh}}_{3}{\mathrm{si}}_{7}$},}\ }\href@noop {} {\bibfield
  {journal} {\bibinfo  {journal} {Phys. Rev. X}\ }\textbf {\bibinfo {volume}
  {8}},\ \bibinfo {pages} {041047} (\bibinfo {year} {2018})}\BibitemShut
  {NoStop}%
\bibitem [{\citenamefont {Bardeen}\ \emph {et~al.}(1957)\citenamefont
  {Bardeen}, \citenamefont {Cooper},\ and\ \citenamefont
  {Schrieffer}}]{Bardeen1957}%
  \BibitemOpen
  \bibfield  {author} {\bibinfo {author} {\bibfnamefont {J.}~\bibnamefont
  {Bardeen}}, \bibinfo {author} {\bibfnamefont {L.~N.}\ \bibnamefont {Cooper}},
  \ and\ \bibinfo {author} {\bibfnamefont {J.~R.}\ \bibnamefont {Schrieffer}},\
  }\bibfield  {title} {\enquote {\bibinfo {title} {Theory of
  superconductivity},}\ }\href {\doibase 10.1103/PhysRev.108.1175} {\bibfield
  {journal} {\bibinfo  {journal} {Phys. Rev.}\ }\textbf {\bibinfo {volume}
  {108}},\ \bibinfo {pages} {1175--1204} (\bibinfo {year} {1957})}\BibitemShut
  {NoStop}%
\bibitem [{\citenamefont {Bouquet}\ \emph {et~al.}(2001)\citenamefont
  {Bouquet}, \citenamefont {Fisher}, \citenamefont {Phillips}, \citenamefont
  {Hinks},\ and\ \citenamefont {Jorgensen}}]{Bouquet2001}%
  \BibitemOpen
  \bibfield  {author} {\bibinfo {author} {\bibfnamefont {F.}~\bibnamefont
  {Bouquet}}, \bibinfo {author} {\bibfnamefont {R.~A.}\ \bibnamefont {Fisher}},
  \bibinfo {author} {\bibfnamefont {N.~E.}\ \bibnamefont {Phillips}}, \bibinfo
  {author} {\bibfnamefont {D.~G.}\ \bibnamefont {Hinks}}, \ and\ \bibinfo
  {author} {\bibfnamefont {J.~D.}\ \bibnamefont {Jorgensen}},\ }\bibfield
  {title} {\enquote {\bibinfo {title} {Specific heat of
  ${\mathrm{mg}}^{11}{B}_{2}$: Evidence for a second energy gap},}\ }\href@noop
  {} {\bibfield  {journal} {\bibinfo  {journal} {Phys. Rev. Lett.}\ }\textbf
  {\bibinfo {volume} {87}},\ \bibinfo {pages} {047001} (\bibinfo {year}
  {2001})}\BibitemShut {NoStop}%
\bibitem [{\citenamefont {Huang}\ \emph {et~al.}(2006)\citenamefont {Huang},
  \citenamefont {Lin}, \citenamefont {Sun}, \citenamefont {Lee}, \citenamefont
  {Kim}, \citenamefont {Choi}, \citenamefont {Lee},\ and\ \citenamefont
  {Yang}}]{Huang2006}%
  \BibitemOpen
  \bibfield  {author} {\bibinfo {author} {\bibfnamefont {C.~L.}\ \bibnamefont
  {Huang}}, \bibinfo {author} {\bibfnamefont {J.-Y.}\ \bibnamefont {Lin}},
  \bibinfo {author} {\bibfnamefont {C.~P.}\ \bibnamefont {Sun}}, \bibinfo
  {author} {\bibfnamefont {T.~K.}\ \bibnamefont {Lee}}, \bibinfo {author}
  {\bibfnamefont {J.~D.}\ \bibnamefont {Kim}}, \bibinfo {author} {\bibfnamefont
  {E.~M.}\ \bibnamefont {Choi}}, \bibinfo {author} {\bibfnamefont {S.~I.}\
  \bibnamefont {Lee}}, \ and\ \bibinfo {author} {\bibfnamefont {H.~D.}\
  \bibnamefont {Yang}},\ }\bibfield  {title} {\enquote {\bibinfo {title}
  {Comparative analysis of specific heat of
  ${\mathrm{yni}}_{2}{\mathrm{b}}_{2}\mathrm{C}$ using nodal and two-gap
  models},}\ }\href@noop {} {\bibfield  {journal} {\bibinfo  {journal} {Phys.
  Rev. B}\ }\textbf {\bibinfo {volume} {73}},\ \bibinfo {pages} {012502}
  (\bibinfo {year} {2006})}\BibitemShut {NoStop}%
\bibitem [{\citenamefont {Huang}\ \emph {et~al.}(2007)\citenamefont {Huang},
  \citenamefont {Lin}, \citenamefont {Chang}, \citenamefont {Sun},
  \citenamefont {Shen}, \citenamefont {Chou}, \citenamefont {Berger},
  \citenamefont {Lee},\ and\ \citenamefont {Yang}}]{Huang2007}%
  \BibitemOpen
  \bibfield  {author} {\bibinfo {author} {\bibfnamefont {C.~L.}\ \bibnamefont
  {Huang}}, \bibinfo {author} {\bibfnamefont {J.-Y.}\ \bibnamefont {Lin}},
  \bibinfo {author} {\bibfnamefont {Y.~T.}\ \bibnamefont {Chang}}, \bibinfo
  {author} {\bibfnamefont {C.~P.}\ \bibnamefont {Sun}}, \bibinfo {author}
  {\bibfnamefont {H.~Y.}\ \bibnamefont {Shen}}, \bibinfo {author}
  {\bibfnamefont {C.~C.}\ \bibnamefont {Chou}}, \bibinfo {author}
  {\bibfnamefont {H.}~\bibnamefont {Berger}}, \bibinfo {author} {\bibfnamefont
  {T.~K.}\ \bibnamefont {Lee}}, \ and\ \bibinfo {author} {\bibfnamefont
  {H.~D.}\ \bibnamefont {Yang}},\ }\bibfield  {title} {\enquote {\bibinfo
  {title} {Experimental evidence for a two-gap structure of superconducting
  $\mathrm{Nb}{\mathrm{se}}_{2}$: A specific-heat study in external magnetic
  fields},}\ }\href@noop {} {\bibfield  {journal} {\bibinfo  {journal} {Phys.
  Rev. B}\ }\textbf {\bibinfo {volume} {76}},\ \bibinfo {pages} {212504}
  (\bibinfo {year} {2007})}\BibitemShut {NoStop}%
\bibitem [{\citenamefont {Chen}\ \emph {et~al.}(2013)\citenamefont {Chen},
  \citenamefont {Jiao}, \citenamefont {Zhang}, \citenamefont {Chen},
  \citenamefont {Yang}, \citenamefont {Nicklas}, \citenamefont {Steglich},\
  and\ \citenamefont {Yuan}}]{Chen2013}%
  \BibitemOpen
  \bibfield  {author} {\bibinfo {author} {\bibfnamefont {J}~\bibnamefont
  {Chen}}, \bibinfo {author} {\bibfnamefont {L}~\bibnamefont {Jiao}}, \bibinfo
  {author} {\bibfnamefont {J~L}\ \bibnamefont {Zhang}}, \bibinfo {author}
  {\bibfnamefont {Y}~\bibnamefont {Chen}}, \bibinfo {author} {\bibfnamefont
  {L}~\bibnamefont {Yang}}, \bibinfo {author} {\bibfnamefont {M}~\bibnamefont
  {Nicklas}}, \bibinfo {author} {\bibfnamefont {F}~\bibnamefont {Steglich}}, \
  and\ \bibinfo {author} {\bibfnamefont {H~Q}\ \bibnamefont {Yuan}},\
  }\bibfield  {title} {\enquote {\bibinfo {title} {Evidence for two-gap
  superconductivity in the non-centrosymmetric compound lanic2},}\ }\href@noop
  {} {\bibfield  {journal} {\bibinfo  {journal} {New Journal of Physics}\
  }\textbf {\bibinfo {volume} {15}},\ \bibinfo {pages} {053005} (\bibinfo
  {year} {2013})}\BibitemShut {NoStop}%
\bibitem [{\citenamefont {Chen}\ \emph {et~al.}(2017)\citenamefont {Chen},
  \citenamefont {Sun}, \citenamefont {Yamada}, \citenamefont {Pyon},\ and\
  \citenamefont {Tamegai}}]{Chen2017}%
  \BibitemOpen
  \bibfield  {author} {\bibinfo {author} {\bibfnamefont {J~T}\ \bibnamefont
  {Chen}}, \bibinfo {author} {\bibfnamefont {Y}~\bibnamefont {Sun}}, \bibinfo
  {author} {\bibfnamefont {T}~\bibnamefont {Yamada}}, \bibinfo {author}
  {\bibfnamefont {S}~\bibnamefont {Pyon}}, \ and\ \bibinfo {author}
  {\bibfnamefont {T}~\bibnamefont {Tamegai}},\ }\bibfield  {title} {\enquote
  {\bibinfo {title} {Two-gap features revealed by specific heat measurements in
  fese},}\ }\href@noop {} {\bibfield  {journal} {\bibinfo  {journal} {Journal
  of Physics: Conference Series}\ }\textbf {\bibinfo {volume} {871}},\ \bibinfo
  {pages} {012016} (\bibinfo {year} {2017})}\BibitemShut {NoStop}%
\bibitem [{\citenamefont {Shang}\ \emph {et~al.}(2020)\citenamefont {Shang},
  \citenamefont {Xie}, \citenamefont {Gawryluk}, \citenamefont {Khasanov},
  \citenamefont {Zhao}, \citenamefont {Medarde}, \citenamefont {Shi},
  \citenamefont {Yuan}, \citenamefont {Pomjakushina},\ and\ \citenamefont
  {Shiroka}}]{Shang2020}%
  \BibitemOpen
  \bibfield  {author} {\bibinfo {author} {\bibfnamefont {T}~\bibnamefont
  {Shang}}, \bibinfo {author} {\bibfnamefont {W}~\bibnamefont {Xie}}, \bibinfo
  {author} {\bibfnamefont {D~J}\ \bibnamefont {Gawryluk}}, \bibinfo {author}
  {\bibfnamefont {R}~\bibnamefont {Khasanov}}, \bibinfo {author} {\bibfnamefont
  {J~Z}\ \bibnamefont {Zhao}}, \bibinfo {author} {\bibfnamefont
  {M}~\bibnamefont {Medarde}}, \bibinfo {author} {\bibfnamefont
  {M}~\bibnamefont {Shi}}, \bibinfo {author} {\bibfnamefont {H~Q}\ \bibnamefont
  {Yuan}}, \bibinfo {author} {\bibfnamefont {E}~\bibnamefont {Pomjakushina}}, \
  and\ \bibinfo {author} {\bibfnamefont {T}~\bibnamefont {Shiroka}},\
  }\bibfield  {title} {\enquote {\bibinfo {title} {Multigap superconductivity
  in the mo5pb2 boron–phosphorus compound},}\ }\href@noop {} {\bibfield
  {journal} {\bibinfo  {journal} {New Journal of Physics}\ }\textbf {\bibinfo
  {volume} {22}},\ \bibinfo {pages} {093016} (\bibinfo {year}
  {2020})}\BibitemShut {NoStop}%
\bibitem [{\citenamefont {Brandt}(1988)}]{brandt1988magnetic}%
  \BibitemOpen
  \bibfield  {author} {\bibinfo {author} {\bibfnamefont {EH}~\bibnamefont
  {Brandt}},\ }\bibfield  {title} {\enquote {\bibinfo {title} {Magnetic field
  density of perfect and imperfect flux line lattices in type ii
  superconductors. i. application of periodic solutions},}\ }\href@noop {}
  {\bibfield  {journal} {\bibinfo  {journal} {J. Low Temp. Phys.}\ }\textbf
  {\bibinfo {volume} {73}},\ \bibinfo {pages} {355--390} (\bibinfo {year}
  {1988})}\BibitemShut {NoStop}%
\bibitem [{\citenamefont {Brandt}(2003)}]{brandt2003properties}%
  \BibitemOpen
  \bibfield  {author} {\bibinfo {author} {\bibfnamefont {Ernst~Helmut}\
  \bibnamefont {Brandt}},\ }\bibfield  {title} {\enquote {\bibinfo {title}
  {Properties of the ideal ginzburg-landau vortex lattice},}\ }\href@noop {}
  {\bibfield  {journal} {\bibinfo  {journal} {Phys. Rev. B}\ }\textbf {\bibinfo
  {volume} {68}},\ \bibinfo {pages} {054506} (\bibinfo {year}
  {2003})}\BibitemShut {NoStop}%
\bibitem [{\citenamefont {Carrington}\ and\ \citenamefont
  {Manzano}(2003)}]{carrington2003magnetic}%
  \BibitemOpen
  \bibfield  {author} {\bibinfo {author} {\bibfnamefont {A}~\bibnamefont
  {Carrington}}\ and\ \bibinfo {author} {\bibfnamefont {F}~\bibnamefont
  {Manzano}},\ }\bibfield  {title} {\enquote {\bibinfo {title} {Magnetic
  penetration depth of mgb2},}\ }\href@noop {} {\bibfield  {journal} {\bibinfo
  {journal} {Phys. C: Supercond.}\ }\textbf {\bibinfo {volume} {385}},\
  \bibinfo {pages} {205--214} (\bibinfo {year} {2003})}\BibitemShut {NoStop}%
\bibitem [{\citenamefont {Damascelli}(2004)}]{Damascelli_2004}%
  \BibitemOpen
  \bibfield  {author} {\bibinfo {author} {\bibfnamefont {Andrea}\ \bibnamefont
  {Damascelli}},\ }\bibfield  {title} {\enquote {\bibinfo {title} {Probing the
  electronic structure of complex systems by arpes},}\ }\href {\doibase
  10.1238/Physica.Topical.109a00061} {\bibfield  {journal} {\bibinfo  {journal}
  {Physica Scripta}\ }\textbf {\bibinfo {volume} {2004}},\ \bibinfo {pages}
  {61} (\bibinfo {year} {2004})}\BibitemShut {NoStop}%
\bibitem [{\citenamefont {McMillan}(1968)}]{PhysRev.167.331}%
  \BibitemOpen
  \bibfield  {author} {\bibinfo {author} {\bibfnamefont {W.~L.}\ \bibnamefont
  {McMillan}},\ }\bibfield  {title} {\enquote {\bibinfo {title} {Transition
  temperature of strong-coupled superconductors},}\ }\href {\doibase
  10.1103/PhysRev.167.331} {\bibfield  {journal} {\bibinfo  {journal} {Phys.
  Rev.}\ }\textbf {\bibinfo {volume} {167}},\ \bibinfo {pages} {331--344}
  (\bibinfo {year} {1968})}\BibitemShut {NoStop}%
\bibitem [{\citenamefont {Allen}\ and\ \citenamefont
  {Dynes}(1975)}]{PhysRevB.12.905}%
  \BibitemOpen
  \bibfield  {author} {\bibinfo {author} {\bibfnamefont {P.~B.}\ \bibnamefont
  {Allen}}\ and\ \bibinfo {author} {\bibfnamefont {R.~C.}\ \bibnamefont
  {Dynes}},\ }\bibfield  {title} {\enquote {\bibinfo {title} {Transition
  temperature of strong-coupled superconductors reanalyzed},}\ }\href {\doibase
  10.1103/PhysRevB.12.905} {\bibfield  {journal} {\bibinfo  {journal} {Phys.
  Rev. B}\ }\textbf {\bibinfo {volume} {12}},\ \bibinfo {pages} {905--922}
  (\bibinfo {year} {1975})}\BibitemShut {NoStop}%
\end{thebibliography}

\end{document}